\documentclass[%
 reprint,
superscriptaddress,
 amsmath,amssymb,
 aps,
 pra,
]{revtex4-1}

\usepackage{graphicx}
\usepackage{dcolumn}
\usepackage{bm}
\usepackage{hyperref}
\usepackage{physics}
\usepackage{xcolor}
\usepackage{enumerate}

\begin{document}

\preprint{APS/123-QED}

\title{Ring-shaped fractional quantum Hall liquids with hard-wall potentials}

\author{E. Macaluso}
\affiliation{%
INO-CNR BEC Center and Dipartimento di Fisica, Universit$\grave{a}$ di Trento, 38123 Povo, Italy
}%
\author{I. Carusotto}%
\affiliation{%
INO-CNR BEC Center and Dipartimento di Fisica, Universit$\grave{a}$ di Trento, 38123 Povo, Italy
}%




\date{\today}

\begin{abstract}
We study the physics of $\nu=1/2$ bosonic fractional quantum Hall droplets confined in a ring-shaped region delimited by two concentric cylindrically symmetric hard-wall potentials. Trial wave functions based on an extension of the Jack polynomial formalism including two different chiral edges are proposed and validated for a wide range of confinement potentials in terms of their excellent overlap with the eigenstates numerically found by exact diagonalization. In the presence of a single repulsive potential centered in the origin, a recursive structure in the many-body spectra and a massively degenerate ground state manifold are found. The addition of a second hard-wall potential confining the fractional quantum Hall droplet from the outside leads to a non-degenerate ground state containing a well defined number of quasiholes at the center and, for suitable potential parameters, to a clear organization of the excitations on the two edges. The utility of this ring-shaped configuration in view of theoretical and experimental studies of subtle aspects of fractional quantum Hall physics is outlined. 
\end{abstract}

\pacs{Valid PACS appear here}
\maketitle


\section{\label{sec:intro}INTRODUCTION}
Since their discovery in low-temperature two-dimensional electron gases subject to strong orthogonal magnetic fields~\cite{Tsui}, fractional quantum Hall (FQH) states have attracted the interest of a constantly increasing number of researchers and have become one of the most active branches of quantum condensed-matter physics. While the interest in this kind of system was initially motivated by their rich variety of astonishing properties~\cite{TongNotes,Goerbig}, more recently it has been further reinforced by long-term perspectives in view of their application in so-called topologically protected quantum computation~\cite{SarmaRMP}.

At the same time, due to recent advances in the field of quantum simulators, FQH physics has drawn the attention also of researchers working on quantum gases of ultracold atoms~\cite{SSLP} and on quantum fluids of light~\cite{ICCC_RMP}. Since they are electrically neutral particles, the orbital motion of both atoms and photons does not experience a Lorentz force in the presence of a magnetic field. As a consequence, observing quantum mechanical orbital magnetic effects in these systems requires the generation of a so-called artificial gauge field.

In the atomic context, an early proposal in this direction was to rapidly rotate trapped atoms and take advantage of the close analogy between the Coriolis force acting on them in the rotating frame and the Lorentz force acting on electrically charged particles~\cite{Cooper}. After that, the idea of associating a Berry phase with the atomic motion by dressing the atoms with suitably designed optical and magnetic fields was exploited~\cite{review_on_synthetic_atoms_Juzeliunas, goldman_juzeliunas}. In recent years, continuous advances in these techniques have led to the nucleation of quantized vortices through synthetic magnetic fields~\cite{Lin:Nature2009} as well as to the observation of many celebrated topological models~\cite{HofstadterColdAtoms,HofstadterKetterle,HaldaneColdAtoms,SpinHallEffectAtoms}.

On the other hand, in the optical framework topologically protected edge states analogous to those displayed in the integer quantum Hall effect have been observed in properly designed magneto-optical photonic crystals~\cite{Wang:Nature2009}, in optical resonator lattices~\cite{hafezi}, and in arrays of waveguides~\cite{rechtsman}. Landau levels for photons, instead, have been recently obtained in non-planar macroscopic ring cavities~\cite{simon}. A detailed review on the recent advances in this field of {\em topological photonics} can be found in~\cite{TOetal_TopologicalPhotonics}.

Whereas the regime of non-interacting particles has already been experimentally investigated extensively in both atomic and optical systems, the realization of strongly correlated states of matter still represent an experimental challenge. While the realization of a fractional quantum Hall state in an atomic gas is made difficult by the still relatively high temperatures achieved in state-or-the-art experiments, the generation of sufficiently large photon-photon interactions and the intrinsic driven-dissipative nature of photonic systems represent the main challenges in the photonic context~\cite{IC_circumnavigating, ICCC_RMP,TOetal_TopologicalPhotonics}. The possibility to engineer strong interactions in the photon gas via coherent dressing of atoms in a Rydberg-EIT configuration was recently studied in~\cite{jia2017rydberg}. Different ways to stabilize fractional quantum Hall states of light through suitable pumping schemes have been investigated in~\cite{Umucalilar:2012PRL,Kapit:2014PRX,OUIC_IncoherentFQH,Mueller_adiabatic_preparation_FQH,SimonFleischhauer_growing_Laughlin_Rydberg}.

From a general perspective, atomic and photonic systems generally offer a wider flexibility in the generation, manipulation, and diagnostics of the quantum state than electronic systems. In particular, one can expect more precise control of the external potential confining the FQH droplet and, as a consequence, of the edge mode properties. In this sense, after years of theoretical studies focused on the properties of harmonically confined FQH droplets~\cite{PhysRevB.60.R16279, PhysRevLett.84.6, PhysRevLett.87.120405, PhysRevLett.91.030402, Cazalilla, PhysRevB.71.121303}, the recent realization of flat-bottomed traps for ultracold atoms~ \cite{Hadzibabic} and the flexibility in designing optical cavities~\cite{simon} and arrays of them~\cite{roushan2016chiral} have motivated researchers to start investigating the effects of hard-wall (HW) confinements on the spectral properties of the excited states of the FQH droplet~\cite{FernSimon,EMIC_disk}. Significant deviations from the standard chiral Luttinger theory for the edge excitations~\cite{Xiao-GangWen}, as well as an extremely accurate one-to-one correspondence between the eigenstates of the FQH cloud and certain trial wave functions expressed in terms of Jack polynomials, were found in~\cite{FernSimon} for idealized HW potentials in the so-called extremely steep limit and then in~\cite{EMIC_disk} for realistic HW potentials.

In this work we provide a detailed study of the physics of a ring-shaped $\nu=1/2$ bosonic FQH droplet confined between two cylindrically symmetric and concentric HW potentials. With respect to the disk-shaped FQH droplets studied in~\cite{EMIC_disk}, ring-shaped droplets are characterized by a much richer variety of excitations due to the presence of two different edges sustaining excitations with opposite chiralities. As the first step, an extension of the Jack polynomial formalism able to properly describe states presenting excitations on both these edges is proposed. The accuracy of this approach is corroborated by the excellent overlap, --typically larger than $98\%$--, of the Jack trial wave functions with the eigenstates found by exact diagonalization calculations. The ground state and the structure of the many-body spectrum of their low-lying excitations are then characterized in detail for two different geometries.

In what we called \emph{pierced geometry}, --i.e., the case in which only the inner repulsive HW potential is present--, a massively degenerate ground state manifold and a recursive organization of the spectra in terms of energy branches and subbranches are observed. The one-to-one correspondence between Jack polynomials and Hamiltonian eigenstates allows us to attribute the observed spectral structures to the presence of specific quasihole and/or edge excitations. In particular, the lowest subbranch of every energy branch consists of a state having $q$ quasiholes and several excitations on its outer edge with little or no energy cost. On the other hand, states belonging to higher energy subbranches are characterized by the presence of excitations on the inner edge and/or a smaller number of quasiholes, which therefore overlap with the inner potential.

The most important results of this work concern the so-called \emph{ring geometry} wherein both the inner and the outer HW potentials are present. In this case, richer many-body spectra are found: The non-degenerate ground state contains a given number of quasiholes at the origin and no edge excitation. Most importantly, a simple way to determine the number of quasiholes in the ground state by looking at the shape of the confining potential in the angular momentum basis with no need for diagonalizing the system Hamiltonian is proposed. The low-lying excited states can be again classified in terms of Jack trial wave functions and parameter regimes displaying a clear structure of edge excitations on the inner and the outer edges are identified. This suggests the possibility of stabilizing photonic FQH states with quasihole excitations by means of a suitable generalization of the pumping scheme proposed in~\cite{OUIC_IncoherentFQH} as well as the possibility of selectively addressing the collective excitations residing on the inner and outer edges.

The structure of the article is the following. In Sec.~\ref{sec:model} we describe the Hamiltonian of the system and we highlight analogies to and differences from our previous work~\cite{EMIC_disk}. In Sec.~\ref{sec:Jacks} we recall the role played by Jack polynomials in the description of the edge and quasihole excitations of disk-shaped FQH droplets and we present a generalization of this formalism for the case of ring-shaped droplets. We study the so-called pierced and ring geometries in Secs.~\ref{sec:pierced_geom} and \ref{sec:ring_geom}, respectively: For both geometries, the excellent overlap of the Jack trial wave functions with the numerical eigenstates found by exact diagonalization is verified and the structure of the many-body spectrum is elucidated. The dependence of the edge state properties on the number of quasiholes pinned in the origin is also discussed. Conclusions are finally drawn in Sec.~\ref{sec:conclusions}.

\section{\label{sec:model}Physical system and theoretical model}

\subsection{Model Hamiltonian}

Similarly to our previous work~\cite{EMIC_disk}, we model the FQH liquid as a two-dimensional system of $\mathcal{N}$ interacting bosons described by a Hamiltonian of the form $H = H_{0} + H_{\mathrm{int}} + H_{\mathrm{conf}}$,
where
\begin{equation}
H_{0} = \sum_{i=1}^{\mathcal{N}} \frac{(\mathbf{p}_{i} + \mathbf{A})^{2}}{2 M}
\end{equation}
is the kinetic energy term in the presence of an orthogonal (synthetic) magnetic field $\mathbf{B} = \nabla \times \mathbf{A} = B \hat{e}_{z}$,
\begin{equation}
H_{\mathrm{int}} = \sum_{i<j} g_{\mathrm{int}} \, \delta^{(2)} (\mathbf{r}_{i} - \mathbf{r}_{j})
\end{equation}
describes contact interactions of strength $g_{\mathrm{int}}$, and
\begin{equation}
H_{\mathrm{conf}} = \sum_{i=1}^{\mathcal{N}} \bigg[ V_{\text{in}} \theta (R_{\text{in}} - | \mathbf{r}_{i} | ) +V_{\text{ext}} \theta (|\mathbf{r}_{i}| - R_{\text{ext}}) \bigg] ,
\label{eq:ring_potential}
\end{equation}
accounts for the spatial confinement of the FQH liquid within the ring-shaped region delimited by a pair of concentric, cylindrically symmetric, repulsive HW potentials: The inner HW radially extends from the origin to $r=R_{\text{in}}$ and has a height $V_{\text{in}}$, while the outer one extends from $r = R_{\text{ext}}$ to infinity with a height $V_{\text{ext}}$ [as usual, $\theta(x)$ denotes the Heaviside step function]. The only addition with respect to our previous work~\cite{EMIC_disk} is the inner HW potential centered at the origin: As we will see in what follows, this apparently minor addition dramatically enriches the physics of the system by creating a spatial hole in the center of the FQH liquid and, consequently, introducing a second {\em inner edge} with opposite chirality.

To take advantage of the underlying cylindrical rotational symmetry of the confining Hamiltonian $H_{\mathrm{conf}}$, we choose to work in the so-called symmetric gauge for the vector potential where $\mathbf{A} = B (-y/2,x/2,0)$. In this gauge, the lowest Landau level (LLL) single-particle wave functions are labeled by the angular momentum $m$ (throughout the article, angular momenta are measured in units of $\hbar$) and have the simple form
\begin{equation}
\varphi_{m} (r,\phi) = \frac{1}{l_{B} \sqrt{2 \pi m!}} e^{i m \phi} \bigg( \frac{r}{\sqrt{2} l_{B}} \bigg)^{m} e^{-r^{2}/4l^{2}_{B}} ,
\label{eq:LLLwfs}
\end{equation}
in terms of the so-called magnetic length $l_{B} = \sqrt{\hbar c /B}$. Since the total angular momentum $L_{z}$ commutes with the Hamiltonian, $[H, L_{z}] = 0$, all many-body eigenstates have a well-defined total angular momentum.

To simplify the theoretical description without losing the many-body physics of the FQH effect, we assume that the magnetic field is strong enough for the cyclotron energy $\hbar \omega_{C} = \hbar B /M$ to be much bigger than all other energy scales set by the characteristic interaction $V_{0} \equiv g_{\mathrm{int}} / 2 l^{2}_{B}$ and potential $V_{\text{in},\text{ext}}$ energies. In this way, we can legitimately make the so-called LLL approximation where the dynamics is restricted to the LLL single-particle states.

Within this approximation, the Hamiltonian can be written in second quantized form in terms of the ladder operators $a^{\dagger}_{m}$ and $a_{m}$, which respectively create and annihilate particles in the LLL state of angular momentum $m$, whose wave function is given by \eqref{eq:LLLwfs}. In particular,
\begin{eqnarray}
H_{0} &=& \varepsilon_{0} \sum_{m} a^{\dagger}_{m} a_{m} ,\\
H_{\mathrm{int}} &=& \frac{g_{\mathrm{int}}}{2 \pi l^{2}_{B}} \sum_{\alpha \beta \gamma \rho} \frac{\Gamma(\alpha + \beta +1)}{\sqrt{\alpha ! \beta ! \gamma ! \rho !}} \frac{ \delta_{(\alpha + \beta , \gamma + \rho)}}{2^{(\alpha + \beta + 2)}} a^{\dagger}_{\alpha} a^{\dagger}_{\beta} a_{\gamma} a_{\rho} ,\\
H_{\mathrm{conf}} &=&  \sum_{m} \mathcal{U}_{m} \, a^{\dagger}_{m} a_{m}= \nonumber \\ &=&\sum_{m} \bigg[ \frac{V_{\text{in}}}{m!}  \gamma_{\downarrow} \bigg( m +1, \frac{R^{2}_{\text{in}}}{2l^{2}_{B}} \bigg) \nonumber \\ && + \frac{V_{\text{ext}}}{m!}  \gamma_{\uparrow} \bigg( m +1, \frac{R^{2}_{\text{ext}}}{2l^{2}_{B}} \bigg) \bigg] a^{\dagger}_{m} a_{m} ,
\label{eq:Hconf}
\end{eqnarray}
where $\varepsilon_{0} = \hbar \omega_{C} / 2$ denotes the kinetic energy of particles in the (widely degenerate) LLL, $\Gamma (t)$ is the Euler $\Gamma$ function
\begin{equation}
\Gamma (t) \equiv \int^{\infty}_{0} x^{t-1} e^{-x} d x ,
\end{equation}
and $\gamma_{\downarrow,\uparrow}(t, R)$ are the so-called lower and upper incomplete $\gamma$ functions
\begin{equation}
\gamma_{\downarrow} (t, R) \equiv \int^{R}_{0} x^{t-1} e^{-x} d x ,
\end{equation}
\begin{equation}
\gamma_{\uparrow} (t, R) \equiv \int^{\infty}_{R} x^{t-1} e^{-x} d x .
\end{equation}
Finally, since all particles in the LLL have the same kinetic energy, within the LLL approximation one can completely neglect the kinetic term and focus on the interaction and confinement Hamiltonian terms, $\tilde{H} = H - H_{0} = H_{\mathrm{conf}} + H_{\mathrm{int}}$.

\subsection{Confining potential}
\begin{figure}[b]
\includegraphics[width=0.4025\textwidth]{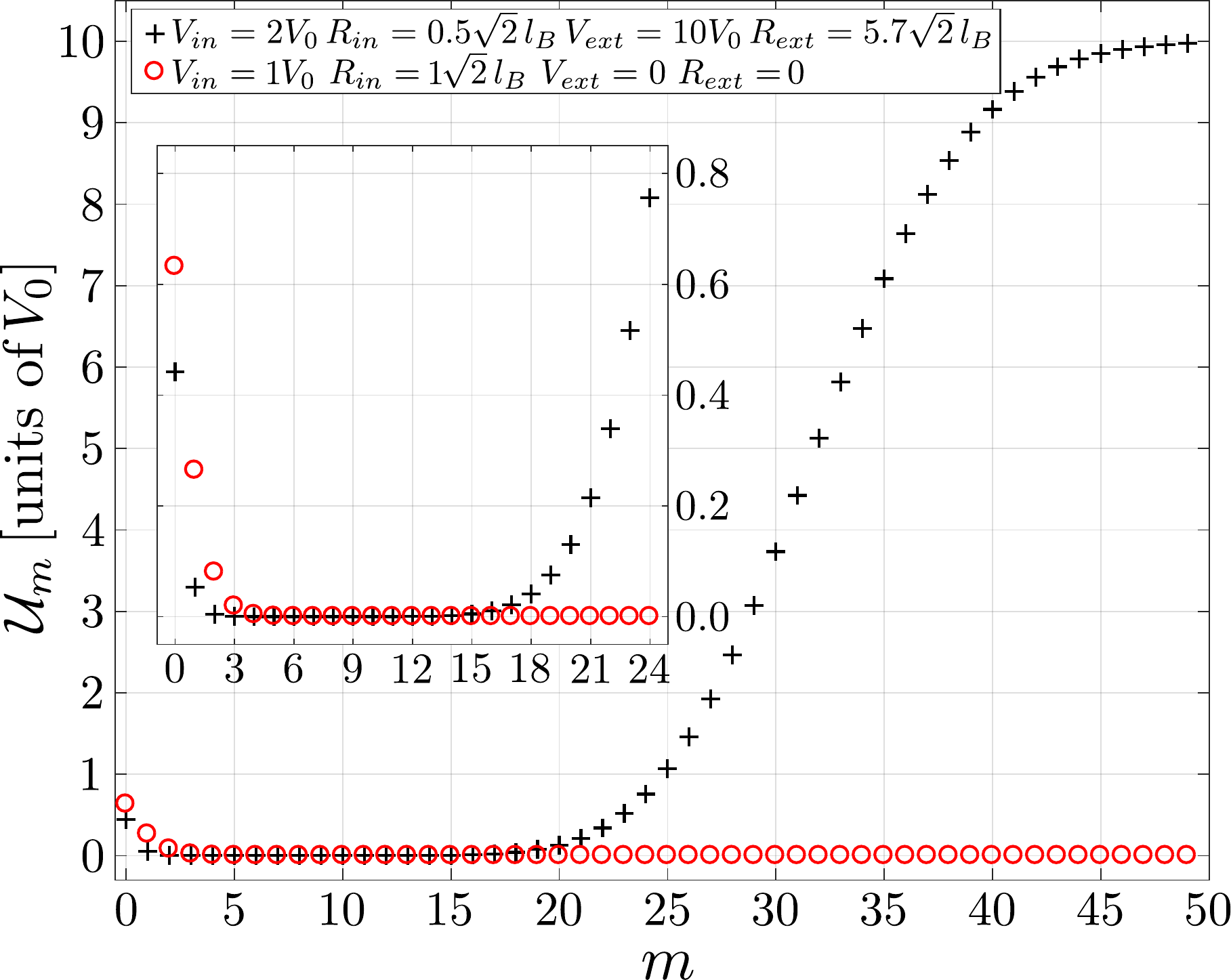}
\caption{Typical behavior of the $\mathcal{U}_{m}$ confinement potentials as a function of the single-particle angular momentum $m$ for two different sets of confining parameters summarized in the legend and corresponding to FQH liquids in, respectively, the ring (black pluses) and pierced (red circles) geometries. The inset provides a magnified view of the main panel. As expected, the extremely steep condition discussed in the text is not fulfilled for the experimentally realistic potential parameters considered here.}
\label{fig:Um_potentials}
\end{figure}

From the confinement Hamiltonian \eqref{eq:Hconf}, it can be immediately seen that, if one restricts the description to the LLL, all information on the confinement potential is encoded in the potential energies $\mathcal{U}_{m}=\langle \varphi_m | H_{\mathrm{conf}} | \varphi_m\rangle$  of the different states of angular momentum $m$. Their typical behavior as a function of $m$ is depicted in Fig.~\ref{fig:Um_potentials} for two different sets of potential parameters $R_{\text{in} (\text{ext})}$ and $V_{\text{in} (\text{ext})}$ which correspond to realistic {\em pierced} and {\em ring} geometries. In the former case, only the inner repulsive HW potential of strength $V_{\text{in}}$ is applied to the atoms and creates a central hole in the FQH liquid but no outer confinement is applied ($V_{\text{ext}}=0$), so the liquid can extend outwards to arbitrary distances. In the latter case, both the inner $V_{\text{in}}$ and the outer $V_{\text{ext}}$ confinements are present, and the liquid is confined within a finite ring-shaped region of space.

As it was discussed in detail in~\cite{EMIC_disk} for the disk geometry with $V_{\text{in}} = 0$, the values of the $\mathcal{U}_{m}$ potentials associated with experimentally realistic confining potentials can hardly satisfy the extremely steep condition $\mathcal{U}_{m-1} \ll \mathcal{U}_{m} \ll \mathcal{U}_{m+1}$ assumed in~\cite{FernSimon}. In Fig.~\ref{fig:Um_potentials}, we show that the same holds for the more general cases considered in this work. For the inner HW potential, the extremely steep condition would in fact read $\mathcal{U}_{m-1} \gg \mathcal{U}_{m} \gg \mathcal{U}_{m+1}$, but the ratio $\mathcal{U}_{m} / \mathcal{U}_{m+1}$ of realistic potentials typically does not exceed values of the order $5$ even at small values of $m$ where, for the values of $R_{\text{in}}$ considered, it is largest. As a result, in the pierced (ring) geometry the confinement energy does not come from a single (two) $\mathcal{U}_{m}$ potential(s), but may receive significant contributions from several $m$ orbitals.

Examples of the many-body spectra of FQH liquids in pierced and ring geometries are displayed in Figs.~\ref{fig:singleHW_spectra} and~\ref{fig:HWHW_spectra} and will be the subject of a detailed discussion later in this work. A crucial tool to unravel the structure underlying their apparent complexity will be the Jack polynomial formalism. In our previous work~\cite{EMIC_disk}, this turned out to be of great utility in the description of the many-body spectra in disk geometries. In Sec.~\ref{sec:Jacks} we start by extending this formalism to the more complex geometries under investigation here. A comparison with numerical exact diagonalization results will then be presented in Secs.~\ref{sec:pierced_geom} and \ref{sec:ring_geom}.

\section{\label{sec:Jacks}Edge States: a Jacks description}

In this section we will recall how Jack polynomials --or simply \emph{Jacks}-- can be used for the analysis of the ground state and the excited states (ESs) of FQH liquids in disk geometries~\cite{EdgeJacks} and will then show how such a formalism can be extended to the ESs of ring-shaped liquids delimited by a pair of edges of opposite chiralities. Such a generalization not only provides trial wave functions with excellent overlap with the exact numerical eigenstates, but also offers an extremely useful tool to understand the structure of the many-body spectrum and study the interplay between quasihole and edge excitations (EEs). Throughout the article we will restrict our attention to ESs lying below the Laughlin gap.

Since the role of Jack polynomials in the FQH context is already widely discussed in the literature, we avoid giving yet another review of the Jack polynomial formalism and rather we refer the interested readers to our previous work \cite{EMIC_disk}, where all the basic concepts needed to fully understand and appreciate the present work are briefly summarized, or to the original papers \cite{PhysRevLett.100.246802, PhysRevB.77.184502, PhysRevLett.103.206801, PhysRevB.84.045127, EdgeJacks}. 

While our Jack polynomial approach can be straightforwardly generalized to generic $\nu = 1/r$ quantum Hall liquids,  from now on, for the sake of simplicity, we restrict ourselves to the specific case of a $\nu = 1/2$ FQH liquid.

\subsection{Excited state wave functions}
In the present FQH context, by ESs we mean all states described by trial wave functions of the form
\begin{equation}
\psi (\{z_{i} \}) \propto S (\{ z_{i} \}) \, \psi_{L} (\{z_{i} \}) ,
\label{eq:edge_wf}
\end{equation}
where
\begin{equation}
\psi_{L}(\{ z_{i}\}) \propto \prod_{i<j} (z_{i} - z_{j})^{2} \, e^{- \sum_{i} |z_{i}|^{2} / 4 l^{2}_{B} } 
\label{eq:Laughlin_wf}
\end{equation}
defines the celebrated $\nu=1/2$ Laughlin wave function and $S (\{ z_{i} \})$ is a generic homogeneous symmetric polynomial in the particle coordinates, whose degree gives the additional angular momentum $\Delta L = L - L_{L}$ on top of the Laughlin state one $L_{L} = \mathcal{N} (\mathcal{N} -1)$~\footnote{We recall that all states of the form \eqref{eq:edge_wf} are characterized by a vanishing interaction energy and that among them the Laughlin state, for which $S (\{ z_{i} \}) = 1$, has the lowest angular momentum $L_L$.}. 

For low additional angular momenta, i.e., $\Delta L \ll \mathcal{N}$, the ESs \eqref{eq:edge_wf} can be interpreted as area-preserving shape deformations of the FQH liquid~\cite{Cazalilla}, which we will call outer edge excitations (OEEs). 
Another particular class of states whose wave functions can be written in this form consist of the quasihole (QH) excitations. While a FQH liquid presenting $n$ QHs at generic positions $\xi_{1}, \dots, \xi_{n}$ can be described by the wave function
\begin{equation}
\psi_{n-\text{QH}} (\{ z_{i}\} ,\{\xi_{i} \}) \propto \bigg( \prod^{\mathcal{N}}_{i=1} \prod^{n}_{j=1} (z_{i} - \xi_{j}) \bigg) \psi_{L}(\{ z_{i} \}),
\end{equation}
in the following we will restrict our attention to the cylindrically symmetric $\xi_{i=1,\ldots n} = 0$ case where all QH's sit at the center and
\begin{equation}
\psi_{n-\text{QH},0} (\{z_{i} \}, \{\xi_{i} = 0 \}) \propto \bigg( \prod_{i} z^{n}_{i} \bigg) \psi_{L} (\{ z_{i} \}).
\label{eq:n-QH_wf}
\end{equation}

\subsection{\label{EEsJacks}Jacks for the edge states}
In general, an ES wave function \eqref{eq:edge_wf} can not be expressed as a single Jack polynomial $J^\alpha_\lambda$ of Jack parameter $\alpha$ and root partition $\lambda$, but can be written as a linear combination of a finite set of them \cite{EdgeJacks}. In particular, any ES with angular momentum $L = L_{L} + \Delta L$ can be written as a linear combination of Jacks having a Jack parameter $\alpha = -2$ and a root partition $\lambda$ that can be obtained by adding~\footnote{The sum of two partitions $\lambda=[\lambda_1,\dots,\lambda_l]$ and $[\mu_1, \dots, \mu_m]$ with $l\geq m$ is defined as a partition of length $l$ whose elements are $\lambda+\mu=[\lambda_1+\mu_1, \dots, \lambda_m+\mu_m,\lambda_{m+1}, \dots, \lambda_l]$. Generalization to $l < m$ cases is straightforward by requiring the commutativity of the partitions sum operation.} to the root partition of the Laughlin state $\Omega  = [2 \mathcal{N} - 2, 2 \mathcal{N} - 4, \dots, 2]$ a generic partition $\eta = [\eta_{1}, \eta_{2}, \dots ]$, --called \emph{edge partition} (EP)--, of the additional angular momentum $\Delta L$, namely,
\begin{equation}
\lambda = \Omega + \eta ,
\label{eq:LaughlinOEEs_root_partition}
\end{equation}
with $|\eta| = \sum_{i} \eta_{i} = \Delta L$.
As the only restriction, for reasons that will be clear in the following, the length of the allowed EPs can not exceed the number of particles $\mathcal{N}$ in the system.

With this prescription for the admissible $\lambda$, it is easy to check that the associated Jack polynomials $J^{\alpha}_{\lambda}$ have total degree equal to $|\lambda| = |\Omega| + |\eta| = L_{L} + \Delta L = L$, and therefore any linear combination of them will be a wave function of total angular momentum $L= L_{L} + \Delta L$, as required. Furthermore, this choice for the possible EPs $\eta$ is in agreement with the degeneracy predicted for the edge states by the chiral Luttinger liquid theory~\cite{Xiao-GangWen}.

Before proceeding, it is worth recalling the biunivocal relation existing between partitions and configurations, that is, many-body states written in the so-called occupation number representation. A given partition $\lambda = [ \lambda_{1} , \dots, \lambda_{N}]$ with $N \leq \mathcal{N}$ can in fact be mapped onto the configuration $\ket{n_{0}(\lambda) \, n_{1}(\lambda) \, n_{2} (\lambda) \, \dots}$ whose occupation numbers $n_{i} (\lambda)$ are given for $i \geq 1$ by the multiplicity of the integer $i$ in the $\lambda$ partition and, for $i=0$, by $n_{0} (\lambda) = \mathcal{N} - N$. As a most remarkable example, the root partition $\Omega = [10, 8, 6, 4, 2]$ of the $\mathcal{N} = 6$ particle $\nu=1/2$ Laughlin state corresponds to the configuration in which the $n_{10} = n_{8}  = n_{6}  = n_{4} = n_{2} = 1$, $n_{0} = 6 - 5 = 1$, and the remaining $n_{i}$ vanish, i.e., $\ket{1\, 0 \, 1 \, 0 \, 1 \, 0 \, 1 \, 0 \, 1 \, 0 \, 1}$.

In terms of configurations, the addition of an EP $\eta$ to the Laughlin root partition $\Omega$ then corresponds to moving particles from the outermost occupied orbitals to even higher $m$ orbitals. To be more precise, given a generic EP $\eta = [\eta_{1}, \eta_{2}, \dots]$, the configuration corresponding to $\lambda = \Omega + \eta$ is the one in which the particle occupying the outermost occupied orbital of the Laughlin root configuration, say, the $m_1$ orbital, is promoted to the $m_1 + \eta_{1}$ orbital, the particle occupying the second outermost occupied orbital $m_2$ is promoted to the $m_2 + \eta_{2}$ orbital, and so on. This connection gives a transparent physical explanation of the restriction that needs to be imposed on the length of the EP $\eta$: No more than $\mathcal{N}$ particles can in fact be moved in an $\mathcal{N}$-particle system.

While Jack polynomials are labeled by their root partition $\lambda$, it is important to keep in mind that they do not coincide with the corresponding configuration. They in fact contain contributions also from those other configurations that are obtained from the root one through all possible {\em squeezing} operations~\cite{PhysRevLett.100.246802,PhysRevB.84.045127}.
In this respect, the recent work~\cite{EdgeJacks} pointed out an interesting relation between the Hilbert spaces spanned by the configurations contributing to different Jacks $J^{\alpha}_{\lambda}$, the so-called \emph{squeezed Hilbert spaces} $\mathcal{H}_{sq}(\lambda)$. They observed that the dominance order $\lambda \succeq \mu$~\footnote{Given two partitions $\lambda$ and $\mu$, we say that $\mu$ is dominated by $\lambda$ or, symbolically, that $\mu \succeq \lambda$ if $\sum^{j}_{i=1} \lambda_{i} \geq \sum^{j}_{i=1} \mu_{i}$ for every $j$.} between two root partitions $\lambda$ and $\mu$ of the same integer $L = L_{L} + \Delta L$ implies the relation
\begin{equation}
\mathcal{H}_{\text{sq}}(\lambda) \subseteq \mathcal{H}_{\text{sq}}(\mu)
\label{eq:squeezed_H_relation}
\end{equation}
between the squeezed Hilbert spaces associated with the two different partitions.

Together with the fact that for any value of the additional angular momentum $\Delta L$ the partition $\lambda_{\Delta L} = \Omega + [\Delta L]$ dominates all the other admissible root partitions obtained as $\Omega + \eta$ with $|\eta| = \Delta L$~\cite{EdgeJacks}, this observation immediately implies that all possible ESs \eqref{eq:edge_wf} of additional angular momentum $\Delta L$ exist in the squeezed Hilbert space $\mathcal{H}_{\text{sq}} (\lambda_{\Delta L})$. Note that the root configuration associated with $\lambda_{\Delta L}$ can be obtained from the Laughlin one by only moving the particle in the outermost occupied orbital $m$ to the $m + \Delta L$ orbital.

Initially, the interest in Jack polynomials in the FQH context was mostly motivated by this last result, which was widely used as a tool in numerical calculations \cite{EdgeJacks, EdgeJacks_densities}. Later on, it was pointed out in \cite{FernSimon,EMIC_disk} that the low energy eigenstates (that is, the ones below the Laughlin gap) of a disk-shaped $\nu = 1/r$ FQH liquid confined by a HW potential can be fully interpreted in terms of the Jacks and the EPs presented above. As we will see in the following of this work, an analogous interpretation in terms of Jacks is also possible for FQH liquids confined by generic potentials of the form \eqref{eq:ring_potential}.

\subsection{Jacks for the quasihole states}
In the preceding section we discussed the general formalism for the description of FQH wave functions \eqref{eq:edge_wf} in terms of Jacks. Although all ingredients to extend the formalism to excitations existing on different edges of a ring-shaped FQH liquid are implicitly present there, the consequences of dealing with such a shape are worth illustrating in more detail. In addition to the presence of two different edges with opposite chiralities, a ring-shaped FQH liquid is in fact characterized by a macroscopic physical hole in the center of its density profile. In FQH language, the central region of vanishing density can be interpreted as due to a certain number of QH excitations sitting in the hole. As a consequence, the starting point for the application of the Jack polynomial formalism to the inner edge excitations (IEEs) of a ring-shaped system cannot be the Laughlin state, but must be an $n$-QH state of the form \eqref{eq:n-QH_wf}, where all $n$ QHs are placed at the origin.

This wave function is of the form \eqref{eq:edge_wf} and has an angular momentum $L_{n-\text{QH}} = L_{L} + n \mathcal{N}$, so it will belong to the $\mathcal{H}_{\text{sq}}(\lambda_{\Delta L = n \mathcal{N}})$ squeezed Hilbert space. Whereas the $n$-QH wave functions correspond in general to linear combinations of Jacks, the case where all $n$ QHs are placed at the origin is a special one \cite{PhysRevLett.101.246806} as its wave function can be described by a single Jack of root configuration $\ket{0^{n} \, 1 \, 0 \, 1  \dots 1 \, 0 \, 1}$.

This state can be obtained from the Laughlin state by promoting each particle from the $m$ orbital to the $m+n$ one. Correspondingly, its root partition can be written as
\begin{equation}
\lambda = \Omega + \kappa^{n},
\label{eq:QHs_root_partition}
\end{equation}
where the \emph{n-\text{QH} partition} $\kappa^{n} \equiv [n, n, \dots, n]$ repeats the integer $n$ for $\mathcal{N}$ times. In this partition, the additional angular momentum $\Delta L = n \mathcal{N}$ is equally distributed among the $\mathcal{N}$ particles.

\subsection{\label{subsec:JacksIEEs}Jacks for the inner edge excitations}

\begin{figure*}[t]
\includegraphics[width=1.0\textwidth]{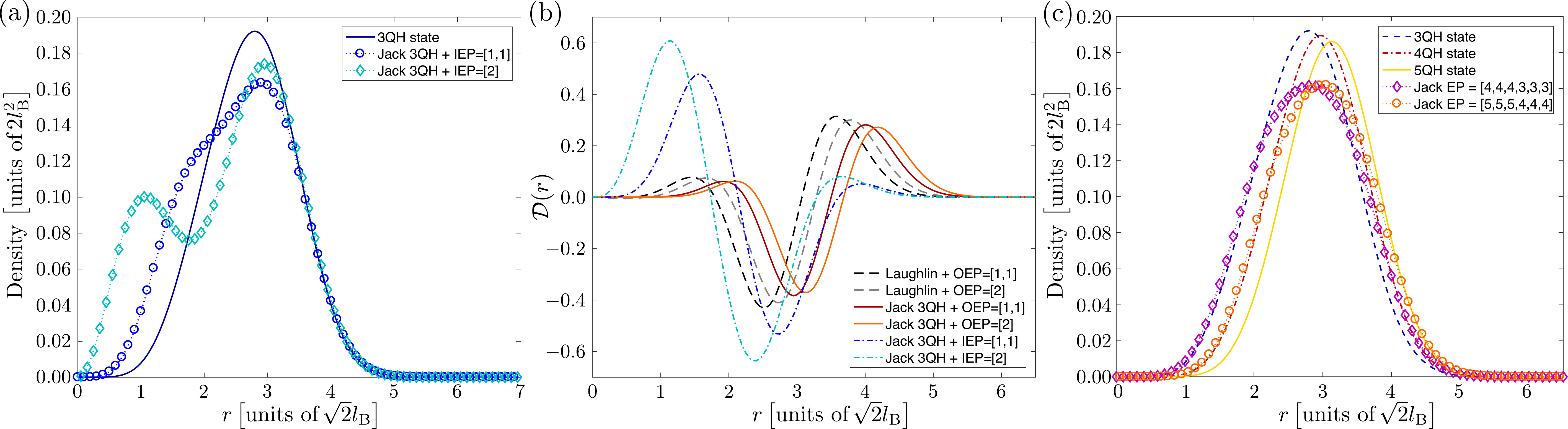}
\caption{(a) Radial density profiles of a $3$-QH state and of Jacks with root partitions $\lambda = \Omega + \kappa^{3} - [1,1] = [13,11,9,7,4,2]$ and $\lambda = \Omega + \kappa^{3} - [2] = [13,11,9,7,5,1]$ for an $\mathcal{N} = 6$ particle system. All density profiles show the same behavior around the outer edge. On the other hand, the density profiles are characterized by very different shapes on the inner edge, as expected from states presenting IEEs. (b) Rescaled density difference $\mathcal{D} (r)$ of different Jacks compared to an $(n=3)$-QH state. The comparison of the $\lambda = \Omega + \rho$ and $\lambda = \Omega + \kappa^{3} + \rho$ cases allows us to identify the latter ones with OEEs of the $3$-QH state: The rescaled density differences only differ by a global shift which reflects the presence of three empty orbitals in the $3$-QH state and a weak attenuation of the peaks, which is consequence of the slightly different length of the edges. At the same time, since the behavior of the $\mathcal{D}(r)$ associated with Jacks of partitions $\lambda = \Omega + \kappa^{3} - \rho$ is similar to those characterizing OEEs of the $3$-QH state, with the difference that the peaks occurs in proximity to the inner edge of the $3$-QH state, we can state that Jacks with partitions of that form well describe the IEEs of the $3$-QH state. Again the difference in the peaks high can be imputed to a difference in the length of the edges. (c) Radial density profiles characterizing the $3$-QH, $4$-QH and $5$-QH states and Jacks with root partitions $\lambda = \Omega + [4,4,4,3,3,3]$ and $\lambda = \Omega + [5,5,5,4,4,4]$. While at small distances the density profile of the former (latter) Jack polynomial is very similar to the one of the $3$-QH state ($4$-QH state), at large distances it follows the behavior of the $4$-QH state ($5$-QH state) density profile.}
\label{fig:IEE-Jack-densities}
\end{figure*}

Excitations of different edges of a ring-shaped FQH liquid are characterized by opposite chiralities. Therefore, since exciting the outer edge induces an increase of the total angular momentum, excitations of the inner edge will lead to a decrease of it.
As a consequence, only states with total angular momenta $L > L_{L}$ can sustain excitations of this kind, which is in agreement with the previous statement that the ground state of a ring-shaped FQH liquid will be reasonably described by wave functions of the form \eqref{eq:n-QH_wf} having total angular momentum $L_{n-\text{QH}} = L_{L} + n \mathcal{N} > L_{L}$.

At this point the extension of Jack formalism to $n$-QH states presenting IEEs is straightforward. Since these states have total angular momentum $L = L_{n-\text{QH}} - \Delta I$, $\Delta I$ being the angular momentum associated with the IEE, in analogy with the discussion of the outer edge in Sec.~\ref{EEsJacks}, one can think that an $n$-QH state presenting an IEE of momentum $\Delta I$ can be obtained as a linear combination of Jacks having as the Jack parameter $\alpha =-2$ and as root partitions all those obtained by subtracting~\footnote{The subtraction of two partitions $\lambda=[\lambda_1,\dots,\lambda_l]$ and $[\mu_1, \dots , \mu_m]$ with $l\geq m$ and $\lambda_{l} \geq \mu_{1}$ is defined as a partition of length $l$ whose elements are $\lambda-\mu=[\lambda_1, \dots, \lambda_{l-m},\lambda_{l-m+1} - \mu_{m}, \dots, \lambda_l- \mu_{1}]$.} from the root partition of the $n$-QH state $\Omega + \kappa^{n}$ any partition $\rho = [\rho_{1}, \rho_{2}, \dots]$, --which we will call \emph{inner edge partition} (IEP)--, of the IEE angular momentum $\Delta I$, namely,
\begin{equation}
\lambda = \Omega + \kappa^{n} - \rho.
\label{eq:IEEs_root_partition}
\end{equation}
In terms of configurations, this corresponds to an inward shift of the particles occupying the lowest $m$ orbitals of the $n$-QH state root configuration $\ket{0^{n} \, 1 \, 0 \, 1  \dots 1 \, 0 \, 1}$. In addition to the constraint already mentioned for the $\eta$ EPs, whose maximum length is fixed by the number of particles $\mathcal{N}$, the IEPs must satisfy the extra condition that the maximum value of their elements $\rho_{i}$ is fixed by the number of QHs in the state sustaining the IEE, i.e.,
\begin{equation}
\rho_{1} \leq n.
\label{eq:rho_constraint2}
\end{equation} 
Note that, since partitions are defined as ordered sequences of positive integers, the constraint \eqref{eq:rho_constraint2} actually imposes that $\rho_{i} \leq n$ for all $i$.

This fact can be easily understood by looking at the root configuration of the $n$-QH state, which is characterized by having the lowest $n$ orbitals empty. Since LLL wave functions \eqref{eq:LLLwfs} are labeled by non-negative integers $m$, the lowest occupied orbital is the $m=n$ one and therefore by moving inwards a particle occupying such an orbital can reduce the total angular momentum at most by $n$. These limitations on the IEPs $\rho$ imply that, for a given value of $\Delta I$, the number of IEEs of a ring-shaped system depends on the internal hole dimension. In particular, for an $n$-QH state the number of different possible IEEs of angular momentum $\Delta I$ is the one predicted by the edge theory of single-branch chiral bosons only for $\Delta I \leq n$.

Finally, dominance relations between the $\lambda$ in \eqref{eq:IEEs_root_partition} for fixed values of $n$ and $\Delta I$ can be inferred from those of the IEPs $\rho$. In particular, given two IEPs $\rho$ and $\rho'$ of $\Delta I$ satisfying $\rho \succeq \rho'$, the partitions $\lambda = \Omega + \kappa^{n} - \rho$ and $\lambda' = \Omega + \kappa^{n} - \rho'$ are related by $\lambda \succeq \lambda'$ and therefore $\mathcal{H}_{\text{sq}} (\lambda) \subseteq \mathcal{H}_{\text{sq}} (\lambda')$. Hence, for a fixed value of $n$, all IEEs of momentum $\Delta I$ are defined in $\mathcal{H}_{\text{sq}} (\lambda_{-\Delta I})$, where $\lambda_{- \Delta I} \equiv \Omega + \kappa^{n} - [\Delta I]$.

\subsection{\label{subsec:IEEs_densities}Density profiles: a benchmark for the EEs}

To corroborate the fact that linear combinations of Jacks with the Jack parameter $\alpha = -2$ and partition $\lambda$ defined in \eqref{eq:IEEs_root_partition} indeed describe the IEEs of an $n$-QH state, we studied the density profiles of these Jacks. For instance, Fig.~\ref{fig:IEE-Jack-densities}(a) shows the density profiles of the two admissible Jacks for $\mathcal{N} = 6$, $n=3$, and $\Delta I = 2$, i.e., $J^{-2}_{[13,11,9,7,4,2]}$ and $J^{-2}_{[13,11,9,7,5,1]}$. As we can see, whereas at large distances their density profiles are almost indistinguishable from the one of the $3$-QH state, at shorter distances their behaviors are very different from it, as expected for states presenting IEEs. Similar results can be obtained for different values of the parameters $\mathcal{N}$, $n$, and $\Delta I$.

To further reinforce our proof based on the analysis of the density profiles, we compared Jacks obtained by either adding or subtracting the same partition $\rho$ to a given $n$-QH partition. As Jacks with root partitions of the form $\lambda = \Omega + \rho$ can be associated with the OEEs of the Laughlin state, the ones with $\lambda = \Omega + \kappa^{n} + \rho$ describe the same OEEs of the outer edge of the $n$-QH state. Pushing this reasoning further, if Jacks with root partitions like those in \eqref{eq:IEEs_root_partition} really describe $n$-QH states presenting IEEs, one can anticipate that by subtracting the same partition $\rho$ from the $n$-QH root, one should induce a similar deformations of the $n$-QH state but this time located on the inner edge.

To verify this guess, we compared the density differences obtained by subtracting the radial density profile of the $n$-QH state from those associated with Jacks having root partition $\lambda = \Omega + \kappa^{n} \pm \rho$. To better highlight the density variations induced on the edges, we focused on the related quantity $\mathcal{D}(r)$ defined as the density difference multiplied by $2 \pi r$, i.e.,
\begin{equation}
\mathcal{D}(r) \equiv 2 \pi r \, \big[ n_{\Omega + \kappa^{n} \pm \rho} (r) - n_{\Omega + \kappa^{n}} (r) \big] ,
\end{equation}
where $r$ denotes the radial distance and $n_{\Omega + \kappa^{n} \pm \rho} (r)$ and $n_{\Omega + \kappa^{n}} (r)$ represent the radial density profiles associated with $J^{-2}_{\Omega + \kappa^{n} \pm \rho}$ and $J^{-2}_{\Omega + \kappa^{n}}$, respectively.

The main results of our calculations are summarized in Fig.~\ref{fig:IEE-Jack-densities}, where we plot the density profiles of the states described by different Jacks. As we can see in Fig.~\ref{fig:IEE-Jack-densities}(b), the density differences obtained by considering the $n$-QH state and the $J^{-2}_{\Omega + \kappa^{n} + \rho}$ Jacks show the same behavior, modulo a translation which simply reflects the presence of $n$ empty orbitals in the $n$-QH state, of those characterizing the OEEs of the Laughlin state. This confirms the expectation that Jacks with root partitions $\lambda = \Omega + \kappa^{n} + \rho$ can indeed be associated with OEEs of the $n$-QH state. 

Even more interestingly, Jacks with $\lambda = \Omega + \kappa^{n} \pm \rho$ describe similar density deformations taking place on the different edges. This confirms our identification of Jacks $J^{-2}_{\Omega + \kappa^{n} - \rho}$ with the IEEs of the $n$-QH state. As a further check, we have verified that the unbalance between the peaks in the rescaled density differences $\mathcal{D}(r)$ for excitations of the inner and the outer edge is a consequence of the shorter length of the inner edge and decreases when the ring dimensions (that is, $n$ and $\mathcal{N}$) are consistently increased.

A final remark regards the competition between states presenting a different number of QHs and excitations on different edges. To be more precise, one could wonder which is the correct interpretation of a Jack polynomial with root partition
\begin{equation}
\lambda = \Omega + \eta = \Omega + [n, \dots, n, n-\rho, \dots, n-\rho],
\label{eq:IEEsVsOEEs}
\end{equation}
where $\eta$ has $\mathcal{N}$ elements, half of which are equal to $n$ and the remaining to $n-\rho$. This partition can in fact be rewritten in the two equivalent ways
\begin{equation}
\lambda =  \Omega + \kappa^{n-\rho} + [\rho, \dots, \rho]
\end{equation}
or
\begin{equation}
\lambda = \Omega + \kappa^{n} - [\rho, \dots, \rho] ,
\end{equation}
where the length of $[\rho, \dots, \rho]$ is given by $\mathcal{N}/2$. The first form suggests that we are dealing with an $(n-\rho)$-QH state presenting an OEE, while the second form rather suggests an $n$-QH states with an IEE. Since the radial density profile of such a state resembles that of the $(n-\rho)$-QH state at short distances and that of the $n$-QH state at large distances [see Fig.~\ref{fig:IEE-Jack-densities}(c)], the two interpretations seem to be equivalent.

In conclusion, we can state that for a given value of $\Delta I$ the above description of QH states presenting IEEs in terms of Jacks correctly applies if $\Delta I \ll \mathcal{N}/2$. On the other hand, for IEPs $\rho$ having both weight $|\rho|$ and length of the order of $\mathcal{N}/2$ the identification of the associated Jacks $J^{-2}_{\Omega + \kappa^{n} - \rho}$ with states showing IEEs is typically much less accurate.

\begin{figure*}
\includegraphics[width=1\textwidth]{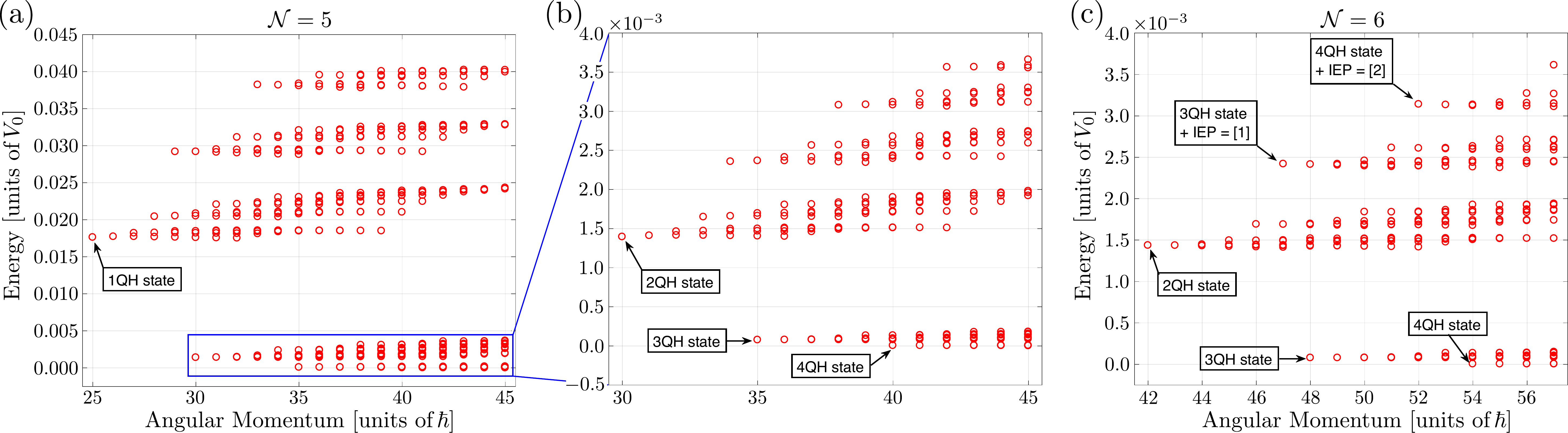}
\caption{Many-body spectra for (a) and (b) $\mathcal{N} = 5$ and (c) $\mathcal{N}=6$ particle systems experiencing a piercing potential of parameters $V_{in} = 2 V_{0}$ and $R_{in} = 0.5 \sqrt{2} l_{B}$. Such spectra clearly show an energetic behavior characterized by the presence of branches and subbranches. By comparing (a) and (b), a recursive structure taking place at different energy scales can also be appreciated. On the other hand, the comparison between (a) and (c) proves that the structure of the spectra does not depend much on the number of particles, as expected from the arguments based on the eigenstate description in terms of Jack polynomials.}
\label{fig:singleHW_spectra}
\end{figure*}

\section{\label{sec:pierced_geom}Pierced Geometry}

After having introduced in the preceding section the Jack polynomial framework, we can now proceed with the study of the many-body energy spectrum for different confinement potentials. In this section we start by considering the case of pierced geometry where the confinement potential \eqref{eq:ring_potential} has $V_{\text{ext}} = 0$ and consists of a single inner repulsive HW potential of strength $V_{\text{in}}$ which extends to the spatial region described by $r < R_{\text{in}}$ and creates a depletion in the FQH liquid density. For this case, many-body spectra showing massively degenerate ground-state manifolds are observed, accompanied by recursive structures of energy branches and subbranches.

The interpretation of these spectral properties in terms of Jacks is quantitatively validated in Sec.~\ref{subsec:Jack_pierced} in terms of the overlap of the Jack trial wave function to the exact diagonalization eigenstates. In the same section, we exploit the possibility of identifying different Jacks with specific excitations of the FQH cloud to give a physical interpretation of the observed spectra in terms of quasihole and edge excitations. A closer study of the inner edge excitations is then discussed in Sec.~\ref{subsec:inner_edge_pierced}, where the existence of an unexpected relation between their density profiles and the value of the angular momentum $\Delta I$ associated with them is highlighted.

\subsection{Jacks description of the energy spectrum}
\label{subsec:Jack_pierced}

Fractional quantum Hall liquids in a pierced geometry are characterized by energy spectra showing a recursive organization in energy branches, which in turn split in a clear and peculiar subbranch structure (see, for instance, Fig.~\ref{fig:singleHW_spectra}). While the qualitative structure of the subbranches is very similar for all branches, its characteristic energy scale is completely different for the different branches. In particular, for a large enough angular momentum $L \geq L_{\mathrm{GSM}}$, we observe a nearly degenerate ground-state manifold (GSM) containing a huge number of states and starting at an angular momentum value $L_{\mathrm{GSM}}$ determined by the potential parameters.

In analogy to the disk-shaped FQH liquid confined by a single HW potential~\cite{EMIC_disk}, we can attempt an interpretation of the many-body spectra in terms of Jacks also for the pierced geometry. As a first step to validate this point of view, we have numerically diagonalized the system Hamiltonian. In particular, for each angular momentum sector, we have considered the Hamiltonian $\tilde{H} = H_{\mathrm{int}}+H_{\mathrm{conf}}$ over an appropriate squeezed Hilbert space (see Appendix~\ref{app:ED}) and we have ensured that the eigenvalues and eigenvectors do not change if we enlarge such a space. The one-to-one correspondence between Jack polynomial trial wave functions and exact diagonalization eigenstates is numerically illustrated in Table~\ref{tab:overlaps}, where overlaps of some numerical eigenstates with the corresponding Jacks are reported. While for Jacks having EPs of the form $\eta = \kappa^{q} + [1, \dots, 1]$ the overlaps can be considered equal to $1$, discrepancies of a few percent are present in the other cases. Despite being very small, these deviations can be almost completely attributed to the slight non-orthogonality of the Jacks, as already found in our previous work~\cite{EMIC_disk}, in which orthogonal trial wave functions of the form $J^{\nu}_{\eta} J^{-2}_{\Omega}$ were found to give better results. Note that for EPs of the form $\eta = \kappa^{q} + [1, \dots, 1]$ the two descriptions coincide, i.e., $J^{-2}_{\Omega+\eta} = J^{\nu}_{\eta} J^{-2}_{\Omega}$, which explains why the overlaps are consistently higher in these cases.

\begin{table}[b]
\begin{tabular}{c | c | c | c}
EP      &    	Eigenstate index	&	$(q,\rho_{1})$	&	$\braket{J^{-2}_{\Omega + \eta}}{\Psi}$ \\
\hline
$\eta = [2,2,2,2,2,2]$			&	$1$st $L=42$						&	$(2,0)$				&	0.9999	\\

\hline
$\eta = [7,2,2,2,2,2]$			&	$1$st $L=47$						&	$(2,0)$				&	0.9886	\\

\hline
$\eta = [6,3,2,2,2,2]$			&	$2$nd $L=47$						&	$(2,0)$				&	0.9710	\\

\hline
$\eta = [5,4,2,2,2,2]$			&	$3$rd $L=47$						&	$(2,0)$				&	0.9584	\\

\hline
$\eta = [5,3,3,2,2,2]$			&	$4$th $L=47$						&	$(2,0)$				&	0.9712	\\

\hline
$\eta = [4,4,3,2,2,2]$			&	$5$th $L=47$						&	$(2,0)$				&	0.9733	\\

\hline
$\eta = [4,3,3,3,2,2]$		 	 &	$6$th $L=47$						&	$(2,0)$				&	0.9795	\\

\hline
$\eta = [3,3,3,3,3,2]$		 	&	$7$th $L=47$						&	$(2,1)$				&	0.9999	\\
\end{tabular}
\caption{Overlaps of some many-body eigenstates $\ket{\Psi}$ calculated by exact diagonalization with the corresponding Jack polynomials for a system of $\mathcal{N} = 6$ particles subject to a piercing potential of parameters $V_{\text{in}} = 2 V_{0}$ and $R_{\text{in}}=0.5 \sqrt{2} l_{B}$. While the Jacks considered are labeled by their EPs, the corresponding ED eigenstates are identified by a couple of integers $(q,\rho_{1})$, corresponding to the branch and subbranch indices, but also through the total angular momentum eigenvalue plus an ordinal number indicating the eigenstate position along the energy axis. For instance, the state denoted by $1$st $L=42$ corresponds to the lowest energy state with $L=42$, that belongs to the $\rho_{1} = 0$ subbranch of the $q=2$ energy branch.}
\label{tab:overlaps}
\end{table}

Specifically, it turns out that the lowest angular momentum state of each branch corresponds to a $q$-QH state, that is, to a Jack polynomial with EP $\kappa^{q} = [q,q, \dots, q]$, while the other states of the branch can be associated with Jack polynomials having EPs of the form $\eta = [\eta_{1}, \eta_{2}, \dots, \eta_{\mathcal{N}-1}, \eta_{\mathcal{N}}=q]$~\footnote{Note that since $\eta$ must be a partition, the condition $\eta_{\mathcal{N}}=q$ implies $\eta_{i} \geq q$ for all $i=1, \dots, \mathcal{N} -1$.}. So any branch can be univocally labeled by the $\eta_{\mathcal{N}}$ value of the states forming it.

Along the same lines, the different subbranches can be associated with the value of the $\mathcal{N}-1$ element $\eta_{\mathcal{N}-1}$ of the EP, or better by the value of $\rho_{1} \equiv \eta_{\mathcal{N}-1} - \eta_{\mathcal{N}}$. In particular, the lowest energy subbranch of the $q$-th branch, that is, the one starting with the $q$-QH state, is formed by all the EEs described by Jacks having $\eta_{\mathcal{N}-1} = q$ which implies $\rho_{1} = 0$, the first excited energy subbranch contains all EEs with $\eta_{\mathcal{N}-1} = (q+1)$ i.e. $\rho_{1} = 1$, and so on for the following increasing values of $\eta_{\mathcal{N}-1}$. Of course, further branched subbranch structures can be observed inside the different subbranches; for them an analogous interpretation in terms of the $\eta_{\mathcal{N}-2}$ values can be put forward.

At this point, it is worth highlighting that the values of the first elements of the EPs do not affect the energy of the state, at least as long as the system is big enough. This not only explains why what happens to the outer edge does not affect the physics of the system, but also suggests that for large enough value of $\mathcal{N}$ the energetic behavior of the FQH liquid in the pierced geometry is almost completely independent of the particle number, as one can see by comparing the many-body spectra in Figs.~\ref{fig:singleHW_spectra}(b) and~\ref{fig:singleHW_spectra}(c).

In the light of this discussion, it is immediate to impute the particular energy scale associated with a given energy branch to the value of $\mathcal{U}_{m}$ for the innermost occupied orbital with $m = \eta_{\mathcal{N}} \equiv q$. Since the $\mathcal{U}_{m}$ values rapidly decrease as a function of $m$ (see the red circles in Fig.~\ref{fig:Um_potentials}), this statement also explains the completely different energy scales observed for consecutive branches. On the other hand, the GSM characterizing this geometry ends up being formed by all states presenting at least $q_{\text{GSM}}$ QHs and their EEs having $\eta_{\mathcal{N}} \geq q_{\text{GSM}}$, which means that $L_{\mathrm{\text{GSM}}} = L_{L} + q_{\text{GSM}} \mathcal{N}$. 

This remarkable feature highlights that a single repulsive HW potential cannot select a non-degenerate ground state presenting a well-defined number of QHs. To obtain this, a suitable external trapping is therefore needed. A detailed discussion of this physics will be the subject of Sec.~\ref{sec:ring_geom} and of Appendix.~\ref{app:HW+harmonic}.

\subsection{A closer look at the inner edge excitations}
\label{subsec:inner_edge_pierced}

Further physically relevant information can be inferred from the previous observations. As we have pointed out in Sec.~\ref{subsec:JacksIEEs}, the set of Jack polynomials with EPs $\eta = [\eta_{1}, \eta_{2}, \dots, \eta_{\mathcal{N}-1}, q]$ also contains states that corresponds to IEEs of QH states containing more than $q$ QHs. To be more precise, since the $k$-th subbranch of the $q$-th branch consists of all states whose EPs have the form $\eta = [\eta_{1}, \eta_{2}, \dots, \eta_{\mathcal{N}-2},(q+k), q]$, the lowest angular momentum state among them is the one with $\eta = [(q+k), \dots, (q+k),q]$. On the other hand, such a partition can be rewritten as $\kappa^{(q+k)} - \rho = [(q+k), \dots, (q+k),(q+k)] - [k]$, which actually coincides with the $(q+k)$-QH state sustaining the IEE described by $\rho = [k]$. So the $k$-th subbranch is composed of such a state and all its OEEs fulfilling the conditions $\eta_{\mathcal{N}-1} = q+k$ and $\eta_{\mathcal{N}}=q$. This observation justifies our choice of using $\rho_{1} \equiv \eta_{\mathcal{N}-1} - \eta_{\mathcal{N}}$ as a label for the different subbranches.

\begin{figure}[b]
\includegraphics[width=0.4025\textwidth]{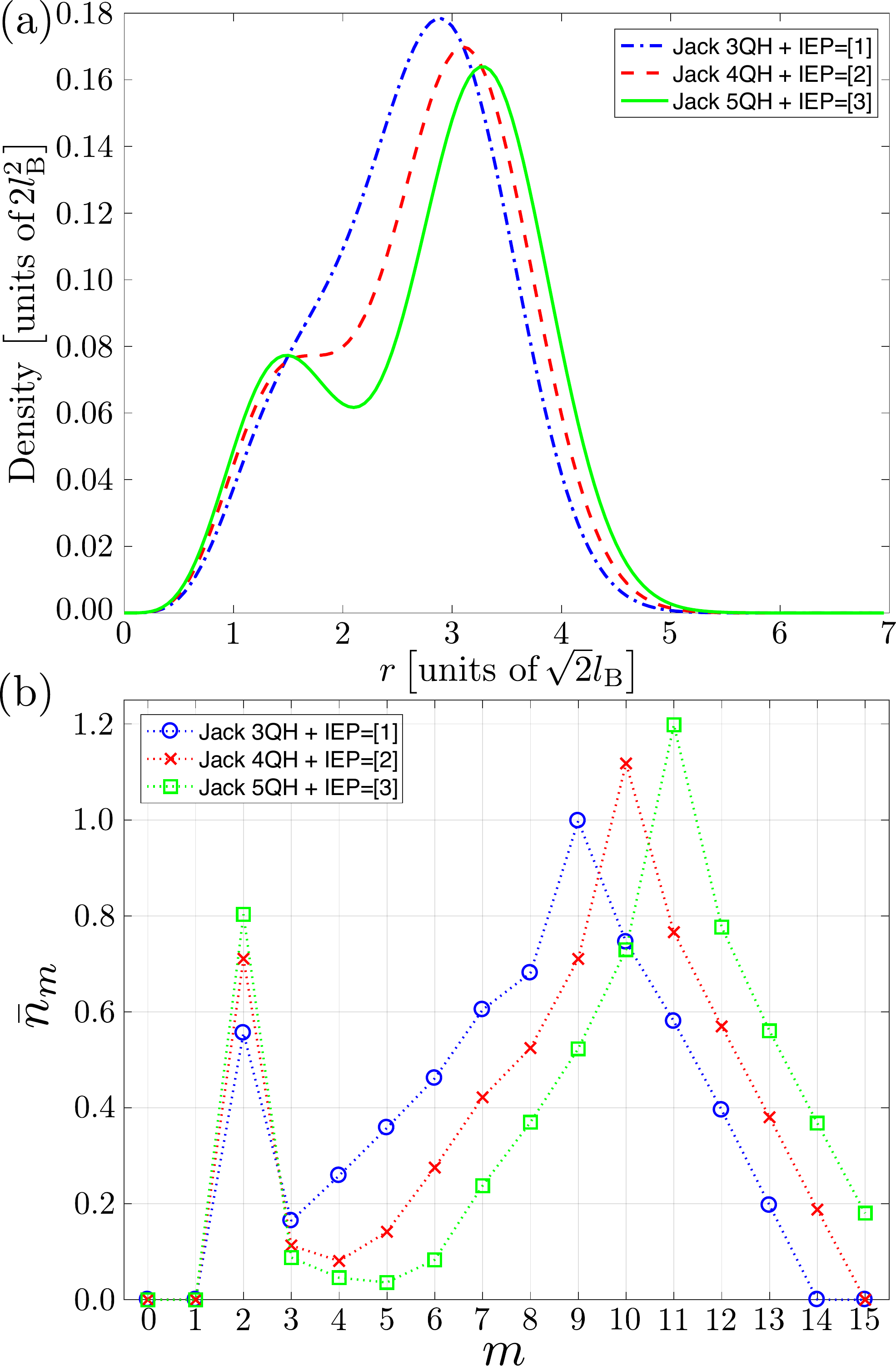}
\caption{(a) Radial density profiles and (b) occupation number expectation values associated with the $3$-QH state with the $\rho = [1]$ IEE, the $4$-QH state with the $\rho = [2]$ IEE, and the $5$-QH state with the $\rho = [3]$ IEE. Despite that for all these states the lowest $m$ occupied orbital is the $m=2$ one, the occupation of this orbital is different for the considered states; in particular, the $5$-QH state with the $\rho = [3]$ IEE is the one characterized by the highest occupation. In terms of densities, this is reflected in the higher density of the $5$-QH state with the $\rho = [3]$ IEE in the inner region $r < 1.5 \sqrt{2} l_{B}$.}
\label{fig:IEEs_occnum_dens}
\end{figure}

\begin{figure*}
\includegraphics[width=1\textwidth]{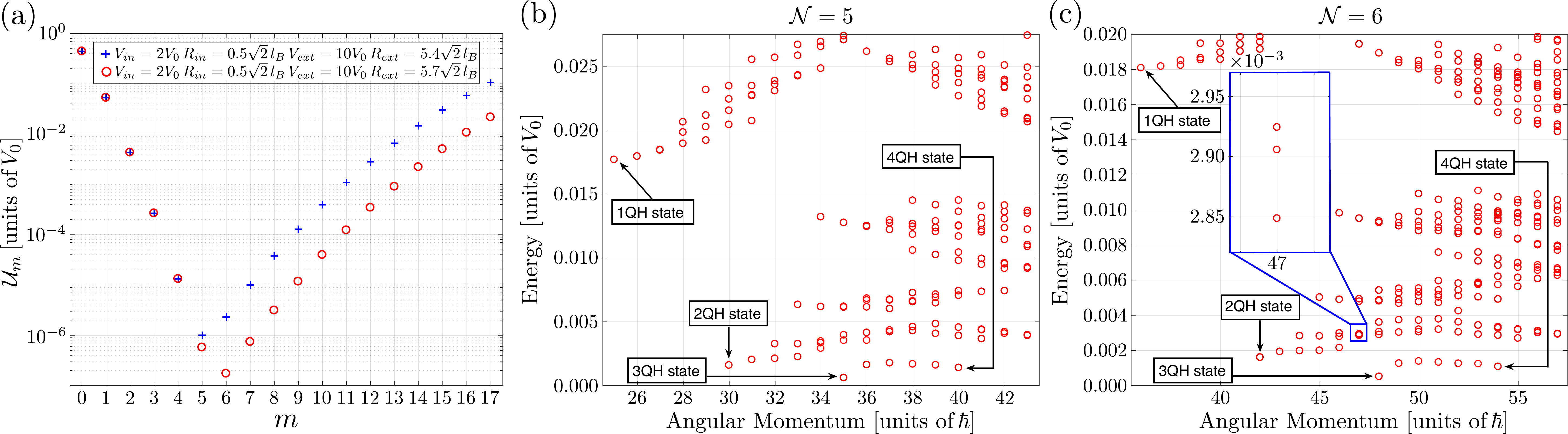}
\caption{(a) Values of the $\mathcal{U}_{m}$ potentials for confinement potentials with $V_{\text{in}} = 2 V_{0}$, $R_{\text{in}} = 0.5 \sqrt{2} l_{B}$, $V_{\text{ext}} = 10 V_{0}$ and $R_{\text{ext}} = 5.4 \sqrt{2} l_{B}$ (blue pluses) or $R_{\text{ext}} = 5.7 \sqrt{2} l_{B}$ (red circles). (b) Many-body energy spectrum of an $\mathcal{N} =5$ particle system subject to the potential illustrated with blue pluses in (a): The ground state is non-degenerate and corresponds to the $3$-QH state, while it is hard to point out any structure in the excited states. (c) Energy spectrum of an $\mathcal{N} =6$ particle system subject to the potential illustrated by red circles in (a). The similarity to the one in (b) is a consequence of the similarity of the $\mathcal{U}_{m}$ in the two cases. In particular, for large $m$ the $U_{m}$ potentials have a very similar shape except for a outward shift by two orbitals. As a result, the energy shifts coming from the outermost and innermost occupied orbitals are roughly the same for Jacks with the same EP describing $\mathcal{N} = 5,6$ particle states. The inset shows the three lowest-energy states in the $L = 47$ angular momentum sector.}
\label{fig:HWHW_spectra}
\end{figure*}

Since the energy of the different states is determined by the overlap with the inner repulsive potential, the above reasoning suggests that the $\rho = [k]$ IEE of the $(q+k)$-QH state (which belongs to a higher energy subbranch) explores the inner regions more substantially than the $\rho = [k-1]$ IEE of the $(q+k-1)$-QH state, which is a quite unexpected fact. To be specific, let us consider as an example the $\rho=[2]$ IEE of the $4$-QH state and the $\rho=[1]$ IEE of the $3$-QH state of an $\mathcal{N}=6$ particle system [see Figs.~\ref{fig:singleHW_spectra}(c) and~\ref{fig:IEEs_occnum_dens}]: The root configuration of the former state is given by $\ket{0 \, 0 \, 1 \, \underline{0} \, 0 \, 0 \, 1 \, 0 \, 1 \, 0 \, 1 \, 0 \, 1 \, 0 \, 1}$, while the root configuration of the latter is $\ket{0 \, 0 \, 1 \, 0 \, 0 \, 1 \, 0 \, 1 \, 0 \, 1 \, 0 \, 1 \, 0 \, 1 \, 0}$. As one can see, the innermost occupied orbital is the same for both root configurations. However, in the first root configuration all five particles occupying the outermost orbitals are shifted by one unit of angular momentum with respect to those of the second root configuration, as emphasized by the presence of an additional empty orbital (underlined in the above formula). Therefore, one could expect that the $4$-QH state sustaining the $\rho=[2]$ IEE would be less affected by the piercing potential as compared with the $3$-QH state sustaining the $\rho=[1]$ IEE. Quite unexpectedly, the plots in Fig.~\ref{fig:IEEs_occnum_dens} actually show that is not the case.

In particular, Fig.~\ref{fig:IEEs_occnum_dens}(a) shows how states belonging to higher energy subbranches are characterized by higher density values in the inner region, in agreement with the naive concept that the $\rho = [k]$ IEE of the $(q+k)$-QH state explores more internal regions with respect to the $\rho = [k-1]$ IEE of the state presenting $q+k-1$ QHs. On the other hand, Fig.~\ref{fig:IEEs_occnum_dens}(b) reports the expectation values of the occupation numbers $\bar{n}_{m} \equiv \expval{a^{\dagger}_{m} a_{m}}{\psi}$ taken on the trial wave functions describing the lowest angular momentum states of the different subbranches: For a given energy branch, the lowest-$m$ occupied orbital is the same for all states since they share the same $\eta_{\mathcal{N}}$ value, but the mean occupation of such an orbital increases (as the energy) as a function of $\rho_{1}$. 

To conclude, even if the pierced geometry will hardly be of experimental interest, its study has taught us several important points. First of all, it has proven that our extension of the Jack polynomial formalism to the two-edge geometry provides an extremely precise description of the Hamiltonian eigenstates and that our identification of some particular Jacks with states presenting IEEs is completely meaningful and physically relevant. In this sense, an unexpected relation was found between IEEs with different $\rho$ and their radial density profiles and occupation number expectation values. Finally, the study of the pierced geometry has highlighted the importance of imposing an additional outer confinement to select a ground state with a well-defined number of quasiholes. This point will be central in the discussion in the next section Sec.~\ref{sec:ring_geom} and in Appendix.~\ref{app:HW+harmonic}.

\section{\label{sec:ring_geom}Ring Geometry}
While in the preceding section we presented the effect of a single repulsive HW potential centered in the origin, we now investigate the effect of the addition of a second HW potential which confines the FQH liquid also from the outside in what we have called \emph{ring geometry}. This section contains the main results of this article. Section~\ref{subsec:GS} is devoted to the discussion of the ground state of the system. In particular, a simple yet efficient method to predict the number of QHs in the ground state is put forward in terms of the confining potential shape in the angular momentum basis. Section~\ref{subsec:energy_branches} then discusses the presence or absence of any regular structure in the energies of the excited states and the possibility of describing them in terms of Jacks. Finally, Sec.~\ref{subsec:QHs&EEs} investigates the interplay between edge and QH excitations, showing that neither the energies nor the density profiles characterizing the edge excitations seem to carry straightforward signatures of the anyonic statistics of quasiholes.

\subsection{\label{subsec:GS}Ground state of the system}
As we can see from Figs.~\ref{fig:HWHW_spectra}(b) and~\ref{fig:HWHW_spectra}(c), FQH liquids subject to a confinement potential with both $V_{\text{in}}, V_{\text{ext}} \neq 0$ are characterized by very complicated energy spectra in which, however, the ground state is clearly non-degenerate. Independently of the particular form of the external potential, this non-degenerate ground state turns out to have an excellent overlap with an $n$-QH state. What depends on the confining parameters is only the number $n$ of QHs. This fact can be easily explained by looking at the shape of the confining potential in the angular momentum space shown in Fig.~\ref{fig:HWHW_spectra}(a) and recalling the general form of the root configurations of the QH states: $\ket{0^{n}\, 1 \, 0\, 1 \, \dots 1 \, 0 \, 1}$. Among all EEs, these root configurations are the densest in angular momentum space, i.e., the sequence of $1$'s and $0$'s does not contain any additional empty orbital. On the other hand, the presence of additional excitations on either one of the two edges implies the presence of extra $0$'s in the root configurations and consequent extra energy contributions coming from higher-valued $\mathcal{U}_{m}$. As a result, both inner and outer EEs of a given $n$-QH state will have a higher energy. Therefore, the ground state of the system must be the $n$-QH state which minimizes the contributions to the energy associated with the different $\mathcal{U}_{m}$.

\begin{figure}[h!]
\includegraphics[width=0.4025\textwidth]{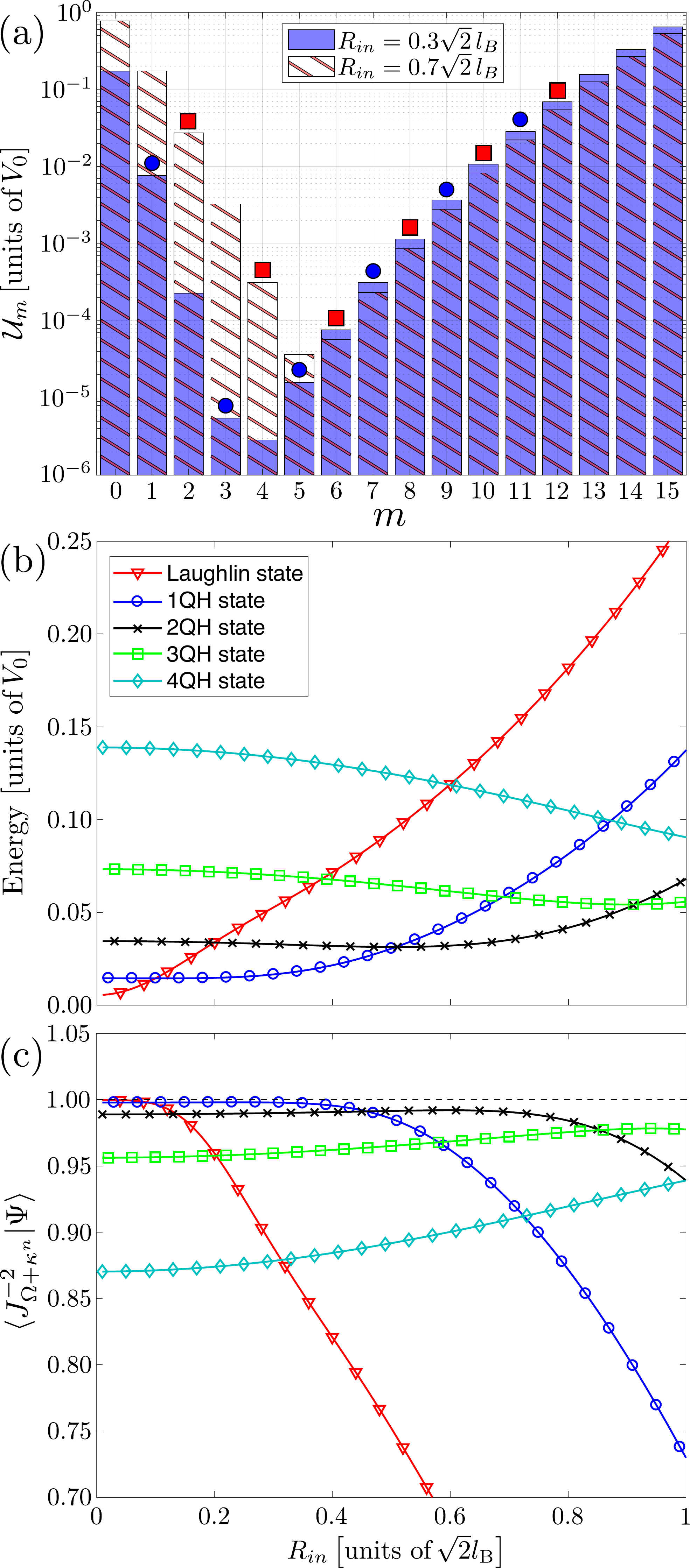}
\caption{(a) The $m$-dependence of the $\mathcal{U}_{m}$ values for $R_{\text{in}} = 0.3 \sqrt{2} l_{B}$ (blue bars) or $R_{\text{in}} = 0.7 \sqrt{2} l_{B}$ (bars with red stripes). Blue circles and red squares summarize the root configuration of the $1$-QH and $2$-QH states, respectively.  (b) Energy of the different $n$-QH states as a function of $R_{\text{in}}$.  As expected, by increasing $R_{\text{in}}$, states containing a higher number $n$ of QHs become energetically favorable. Moreover, by comparing (a) and (b), we can note that the lowest energy QH state is the one whose root configuration minimizes the energy contribution coming from the different $\mathcal{U}_{m}$. (c) Overlap between numerical states $\ket{\Psi}$ and Jacks describing the associated QH states, which proves that for any value of $R_{\text{in}}$ the lowest energy states is extremely well described by an $n$-QH state. For all plots $V_{\text{in}} = 2 V_{0}$, $V_{\text{ext}} = 100 V_{0}$ and $R_{\text{in}}$, $R_{\text{ext}}$ satisfying Eq. \eqref{eq:constant_surface_condition} with $\mathcal{A} = \pi (5.25 \sqrt{2} l_{B})^{2}$ have been considered. }
\label{fig:QHnumber}
\end{figure}

The above argument not only explain why the ground state of a FQH liquid in the ring geometry is always given by an $n$-QH state, but also allows one to predict the number $n$ of QHs in the ground state by simply looking at the $\mathcal{U}_{m}$ values. This statement has been numerically proved by investigating the dependence of the QH number $n$ on the potential parameters (see Fig.~\ref{fig:QHnumber}). Given the massive degeneracy of the ground state manifold discussed in Sec.~\ref{sec:pierced_geom} for the pierced geometry, the number $n$ of QHs in the degenerate ground state  depends not only on the parameters $V_{\text{in}}$ and $R_{\text{in}}$ of the inner piercing potential but also on the ones of the external confinement. So, in order to study the dependence of $n$ on $R_{in}$, for a fixed value of $V_{in}$, we introduced a \emph{constant surface condition} to ensure the balance between the inner and the outer HWs. Such a condition is based on the FQH liquid incompressibility and consists in choosing values of $R_{\text{in}}$ and $R_{\text{ext}}$ which satisfy
\begin{equation}
\pi R^{2}_{\text{ext}} - \pi R^{2}_{\text{in}} = \mathcal{A},
\label{eq:constant_surface_condition}
\end{equation}
where $\mathcal{A}$ is given by the disk-shaped surface delimited by a single HW potential confining the FQH cloud in the weak-confinement regime~\cite{EMIC_disk}, i.e., a HW potential which selects the Laughlin state as the non-degenerate ground state of the system and removes the degeneracy between the different ESs while still keeping many of them below the Laughlin gap.

Within this condition, the energetic behavior of different QH states as a functions of $R_{\text{in}}$ is presented in Fig.~\ref{fig:QHnumber}.
Figure~\ref{fig:QHnumber}(b) shows that states with an increasing number of QHs become energetically favorable as the piercing potential gets wider. At the same time, Fig.~\ref{fig:QHnumber}(c) shows that for intervals of $R_{\text{in}}$ values in which a given state has the lowest energy, its overlap with the corresponding $n$-QH state is higher than $95 \%$. The deviations from total overlap can be attributed to mixing with states lying above the Laughlin gap induced by the confinement. On the other hand, the fact that the overlaps (the energies) decrease (increase) faster when we move from their maxima (minima) towards larger values of $R_{\text{in}}$ than when we move towards smaller ones can be attributed to the very different strengths considered in this plot for the inner and the outer HW potentials.

Finally, Fig.~\ref{fig:QHnumber}(a) displays (in logarithmic scale) the $\mathcal{U}_{m}$ values for the two $R_{\text{in}} = 0.3 \sqrt{2} l_{B}$ and $R_{\text{in}} = 0.7 \sqrt{2} l_{B}$ cases and illustrates a simple way to predict the number of QHs in the ground state for a given external potential. Consider, for instance, the $R_{\text{in}} = 0.7 \sqrt{2} l_{B}$ case: Red squares denotes particles in the many-body state described by the $2$-QH state root configuration, i.e., $\ket{n_{0} \, n_{1} \, n_{2} \dots} = \ket{0 \, 0 \, 1 \, 0 \, 1 \, 0 \, 1 \, 0 \, 1 \, 0 \, 1 \, 0 \, 1}$. Among the root configurations characterizing the different $n$-QH states, this configuration minimizes the energy contribution coming from the $\mathcal{U}_{m}$, in agreement with the QH state energetic behaviors reported in Figure~\ref{fig:QHnumber}(b). Root configurations of the $1$-QH and $3$-QH states would in fact imply the occupation of the $m=1$ and $m=13$ orbitals, respectively, which are both characterized by much-higher-$\mathcal{U}_{m}$ values.

Although a method which allows us to know the system ground state without diagonalizing the Hamiltonian, and with such a high accuracy, is remarkable just by itself, it also represents a very powerful tool to push forward numerical calculation. These naive considerations, together with the set of properties characterizing the so-called squeezed Hilbert spaces, indeed allow us to properly and efficiently construct the Hilbert space for the diagonalization procedure, as explained in Appendix~\ref{app:ED}.

\subsection{\label{subsec:energy_branches}Energy branches and Jacks description}
As compared to the disk geometry discussed in~\cite{EMIC_disk} and to the pierced geometry discussed in Sec.~\ref{sec:pierced_geom}, it is generally more difficult to identify a regular structure in the energy spectra for the ring geometry shown in Fig.~\ref{fig:HWHW_spectra}. Nonetheless, the one-to-one correspondence between exact diagonalization eigenstates and Jack polynomials is perfectly maintained, as one can see in Table~\ref{tab:overlaps_ring}.

It is however important to stress that this difficulty is not intrinsic to the ring geometry but rather stems from the small size of the system we are considering. Dealing with few-particle systems limits in fact the size of the inner piercing potential and, unless careful attention is paid, typically results in very asymmetric shapes of the $\mathcal{U}_{m}$ with respect to their minimum. As a consequence, states presenting the same kind of edge excitations but on different edges (that is, IEEs and OEEs associated with EPs $\eta = \kappa^{n} \pm \rho$) are typically characterized by very different shifts in energy, which underlies the complicate and nested structure of the many-body energy spectra shown in Fig.~\ref{fig:HWHW_spectra}.

\begin{table}[t]
\begin{tabular}{c | c | c }
EP      &    	Eigenstate index	&	$\braket{J^{-2}_{\Omega + \eta}}{\Psi}$ \\
\hline
$\eta = [3,3,3,3,3,2]$			&	$1$st $L=47$						&	0.9954	\\

\hline
$\eta = [4,3,3,3,2,2]$			&	$2$nd $L=47$						&	0.9813	\\

\hline
$\eta = [4,4,3,2,2,2]$			&	$3$rd $L=47$						&	0.9926	\\

\hline
$\eta = [4,4,3,3,3,3]$			&	$1$st $L=50$						&	0.9961	\\

\hline
$\eta = [4,4,4,4,2,2]$			&	$2$nd $L=50$						&	0.9915	\\

\hline
$\eta = [5,3,3,3,3,3]$			&	$3$rd $L=50$						&	0.9978	\\

\hline
$\eta = [4,4,4,3,3,2]$		 	 &	$4$th $L=50$						&	0.9792	\\

\hline
$\eta = [4,4,4,3,3,3]$		 	&	$1$st $L=51$						&	0.9970	\\

\hline
$\eta = [5,4,3,3,3,3]$		 	&	$2$nd $L=51$						&	0.9865	\\

\hline
$\eta = [4,4,4,4,3,2]$		 	&	$3$rd $L=51$						&	0.9811	\\

\hline
$\eta = [6,3,3,3,3,3]$		 	&	$9$th $L=51$						&	0.9933	\\
\end{tabular}
\caption{Overlaps of several Hamiltonian eigenstates $\ket{\Psi}$ with the corresponding Jack polynomials for a system of $\mathcal{N} = 6$ particles experiencing two HWs, respectively of parameters $V_{\text{in}} = 2 V_{0}$ and $R_{\text{in}}=0.5 \sqrt{2} l_{B}$ and parameters $V_{\text{ext}} = 10 V_{0}$ and $R_{\text{ext}}=5.7 \sqrt{2} l_{B}$. As in Table \ref{tab:overlaps}, the Jacks considered are labeled by their EPs $\eta$, while the corresponding ED eigenstates are identified through the total angular momentum eigenvalue plus an ordinal number indicating the eigenstate position along the energy axis.}
\label{tab:overlaps_ring}
\end{table}

\begin{figure}[t]
\includegraphics[width=0.80\columnwidth]{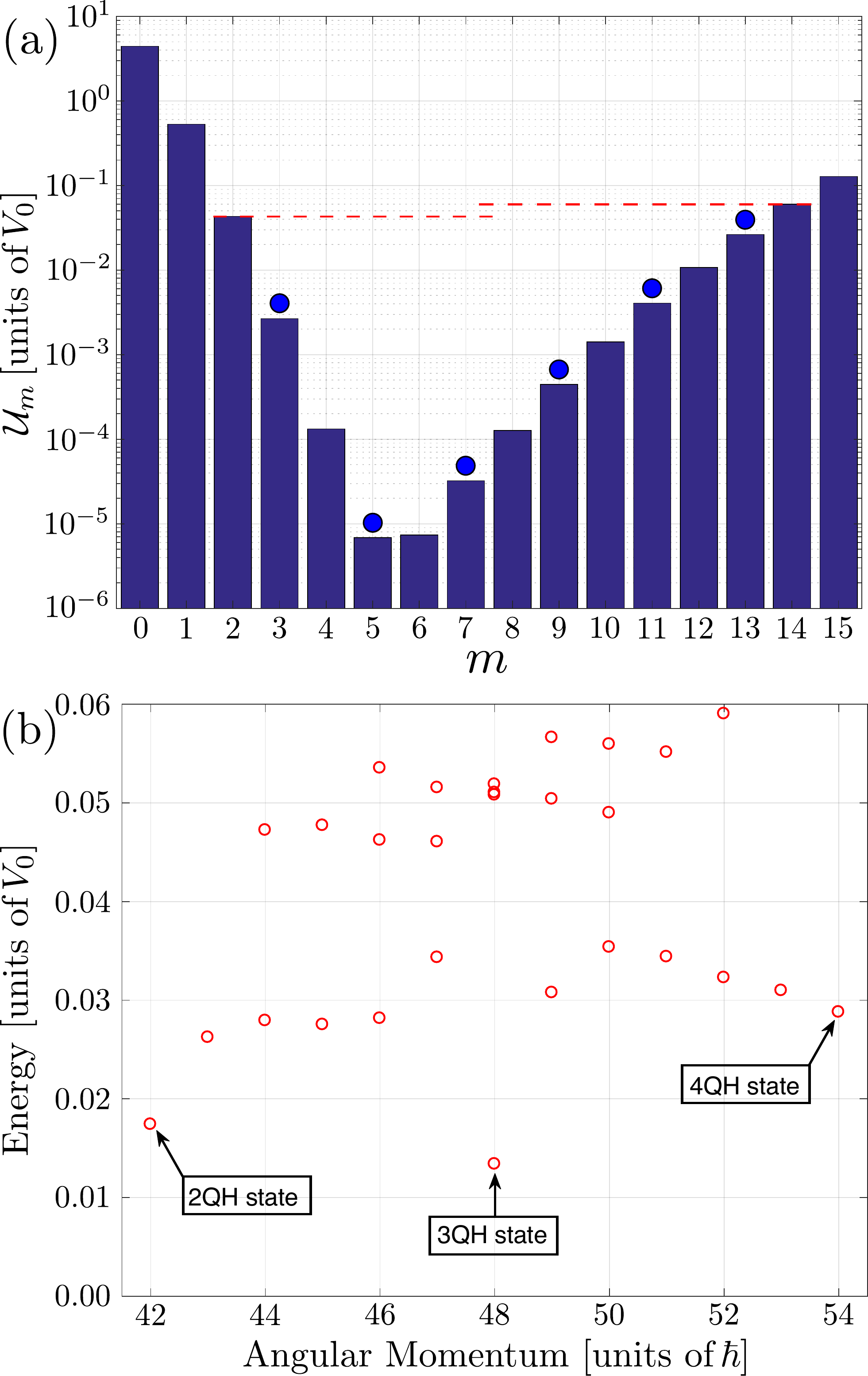}
\caption{(a) Plot of the $\mathcal{U}_{m}$ values for a confinement potential of parameters $V_{\text{in}} = 20 V_{0}$, $R_{\text{in}} = 0.5 \sqrt{2} l_{B}$, $V_{\text{ext}} = 75 V_{0}$ and $R_{\text{ext}} = 5.5 \sqrt{2} l_{B}$. (b) Associated $\mathcal{N} = 6$ particles spectrum. Blue circles in (a) indicate the occupied orbitals in the root configuration of the $3$-QH state. By recalling that root configurations for the IEEs (OEEs) can be obtained by moving inward (outward) particles occupying the innermost (outermost) single-particle orbitals, this representation of the ground state root configuration pictorially explains which orbitals are associated with a given inner or outer edge excitation. Since the $\mathcal{U}_{2}$ and $\mathcal{U}_{14}$ values (red dashed lines) are comparable, the Hamiltonian eigenstates in the spectrum (or at least those with $n_{i} = \expval{a^{\dagger}_{i} a_{i}}{\Psi} = 0$ $\forall i<2$ or $i>14$) are organized in an almost symmetric way. Eigenstates of the lowest branch between the $2$-QH and the $3$-QH state correspond to Jacks with EPs of the form $\eta = [3, \dots, 2]$ and are characterized by a non-vanishing occupation of the $m=2$ orbital, while those between the $3$-QH and the $4$-QH state have non-vanishing occupation of the $m=14$ orbital and are well described by Jacks with $\eta = [4, \dots, 3]$. Finally, states in the upper branches can be associated with Jacks of EPs like $\eta=[4,\dots,2]$ and therefore describing states presenting both inner and outer EEs.}
\label{fig:symmetric_spectrum}
\end{figure}

A clear example of this complexity is represented by the lowest-energy eigenstate of angular momentum $L = 47$, which is almost degenerate with the second and third excited states in the same angular momentum sector [see the inset in Fig.~\ref{fig:HWHW_spectra}(c)]. As it is confirmed by the overlaps in Table~\ref{tab:overlaps_ring}, such a state corresponds to the $\rho = [1]$ IEE of the $3$-QH state, while the second and the third excited states present excitations of both edges with EPs $\eta = [4,3,3,3,2,2]$ and $\eta = [4,4,3,2,2,2]$, respectively. Since they involve also an excitation on the outer edge, these two states differ from the first excited one in a non-vanishing occupation of the $m=14$ single-particle orbital. On the other hand, they are also characterized by a slightly lower occupation of the $m=2$ orbital. Since $\mathcal{U}_{14} < \mathcal{U}_{2}$ [see Fig.~\ref{fig:HWHW_spectra}(a)], this however results into comparable energy shifts to the first excited state, even if the latter only presents excitation on a single edge.

In spite of this overall complexity of the spectrum, a manifest organization of the many-body energy spectra in energy branches can be identified if one carefully chooses the potential parameters so that $\mathcal{U}_{2}$ and $\mathcal{U}_{14}$ have comparable values. In this way, configurations can be obtained in which EEs described by partitions of the form $\eta = \kappa^{n} \pm [1,\dots]$ (we recall that opposite signs are associated with excitation of opposite edges) are almost equally affected by the external confinement. An example of such a configuration is illustrated in Fig.~\ref{fig:symmetric_spectrum}, where we show the many-body spectrum for an $\mathcal{N} = 6$ particle system experiencing a suitably designed confinement of parameters $V_{\text{in}} = 20 V_{0}$, $R_{\text{in}} = 0.5 \sqrt{2} l_{B}$, $V_{\text{ext}} = 7 V_{0}$ and $R_{\text{ext}} = 5.5 \sqrt{2} l_{B}$.

Focussing our attention on the lowest-energy states, we can see here that the ground state of corresponds to the $3$-QH state, in agreement with the previously presented criterion. At the same time, the lowest-energy excitations are symmetrically organized in two distinct energy branches: The lower branch extends from the $2$-QH state to the $4$-QH one and contains states presenting excitations described by $\rho$ of the form $[1, \dots]$ on only one of the two edges of the $3$-QH state. 

The upper branches are formed by states sustaining excitations of this kind on both edges.  Since IEEs with $\eta = \kappa^{3} - [1,\dots]$ and OEEs with $\eta = \kappa^{3} + [1,\dots]$ are respectively characterized by non-vanishing occupations of the $m=2$ and the $m=14$ orbitals, this condition implies that the effect of the confinement on these states is roughly of the same magnitude independently of which edge is actually excited. This simple criterion will be of great utility in designing the confinement potential to be used for bigger systems.

\subsection{\label{subsec:QHs&EEs}Interplay between QHs and EEs}
\begin{figure}[b]
\includegraphics[width=0.475\textwidth]{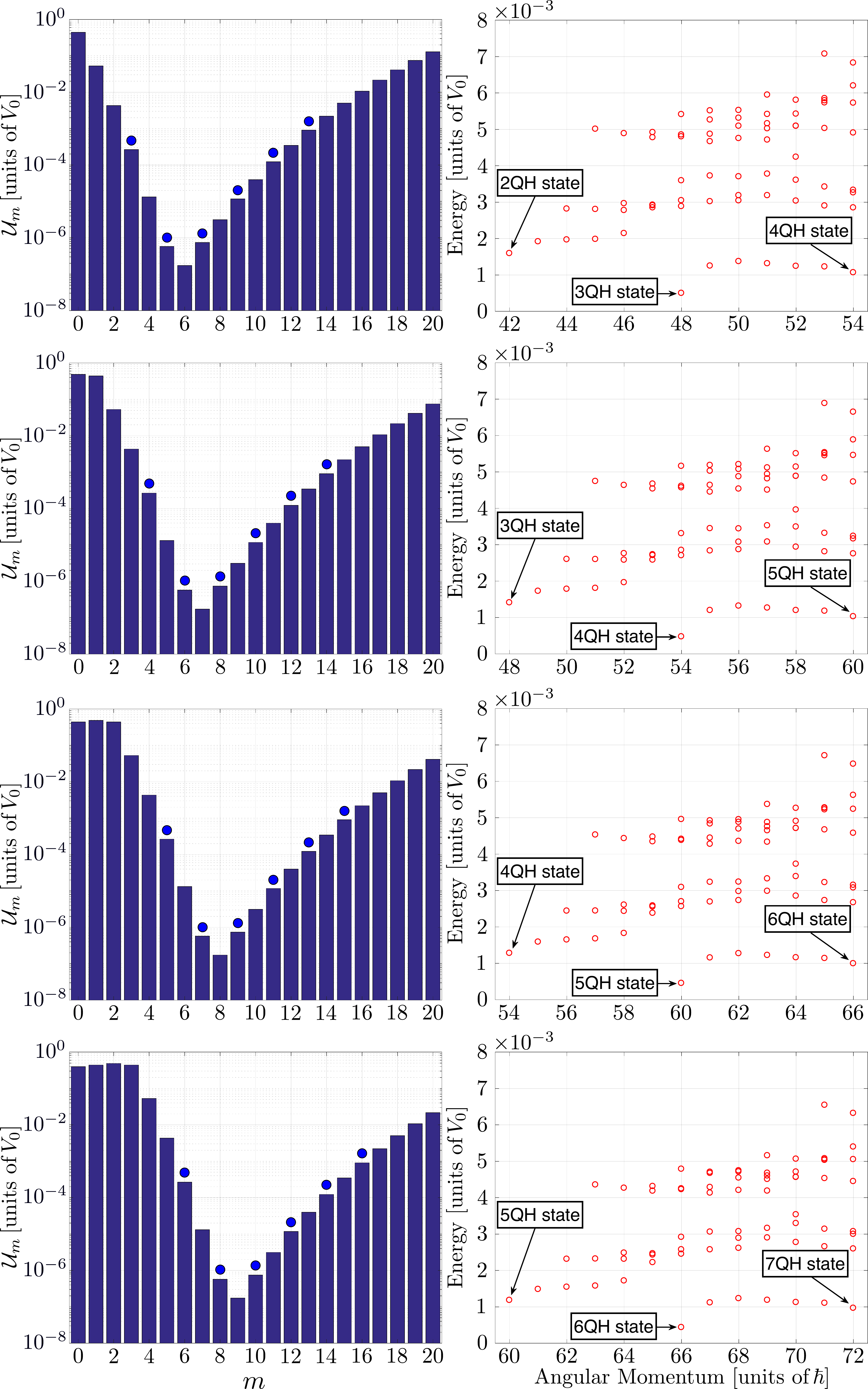}
\caption{In the left column, the top plot shows the $\mathcal{U}_{m}$ potentials associated with the confining parameters $V_{\text{in}} = 2 V_{0}$, $R_{\text{in}} = 0.5 \sqrt{2} l_{B}$, $V_{\text{ext}} = 10 V_{0}$, and $R_{ext} = 5.7 \sqrt{2} l_{B}$ and the three lower plots display mock $\mathcal{U}_{m}$ potentials obtained by shifting the previous ones by one, two and three orbitals, respectively. In each panel, the blue circles indicate again the occupied orbitals in the root configurations of Jacks describing the ground state. The right column shows the corresponding many-body energies. Their structure is almost the same for all cases considered and no periodic behavior can be observed as a function of the number of QHs. The slow decrease of the excited-state energies as a function of the number of QHs pinned in the origin can be imputed to the bigger length of the ring-shaped system edges.}
\label{fig:shifted_pot}
\end{figure}

In this section we briefly discuss how the general theory of FQH liquids in a ring geometry presented in the previous sections may have some utility in addressing intriguing physical questions such as whether the excitations of the edges are affected in any way by the number of QHs around which they are moving or whether one can use the energy spectra or the density profiles associated with the different EEs to probe the anyonic statistics of the QHs. Even though the answer to these questions is at the present level of understanding negative, this result does not diminish the importance and the utility of the formalism we have developed.

\begin{figure}[b]
\includegraphics[width=0.38\textwidth]{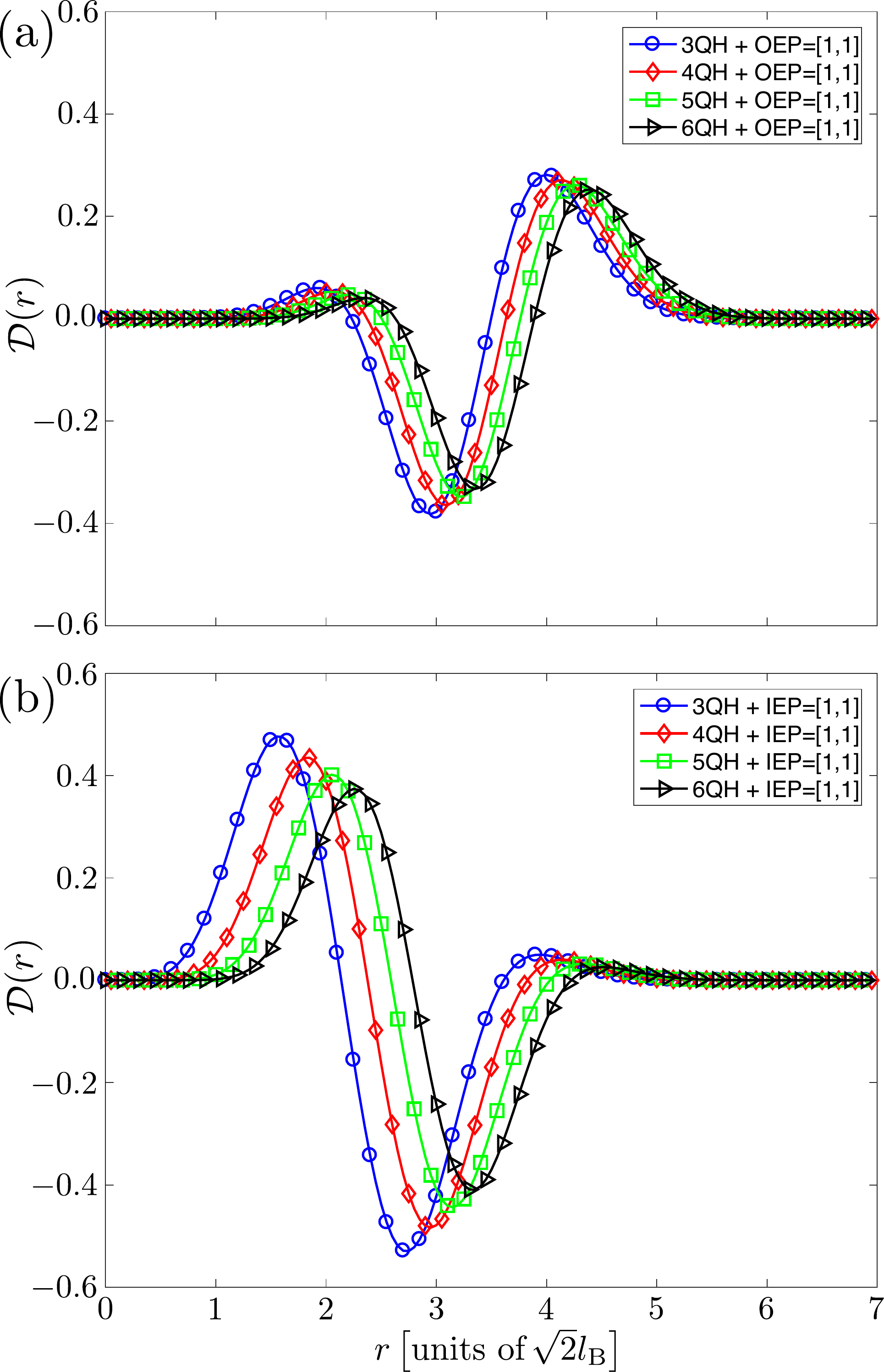}
\caption{Rescaled density differences $\mathcal{D}(r)$ associated with the same (a) OEE or (b) IEE lying on the edges of QH states containing a different number of QHs. To be precise, the OEE (IEE) with OEP (IEP) of the form $[1,1]$ has been considered. As it is clearly visible in both panels, having additional QHs in the origin simply results in a outward translation of the curve describing $\mathcal{D}(r)$, which reflects the presence of additional empty orbitals, and in an attenuation of the oscillations characterizing $\mathcal{D}(r)$, which is the consequence of having a longer edge.}
\label{fig:shifted_dens}
\end{figure}

As a first step in these directions, we have considered different ring geometries, each of them characterized by a ground state containing a different number of QHs. Moreover, to ensure that the only difference between the considered configurations is the number of QHs pinned at the origin, we focus on {\em ad hoc} potentials artificially built in terms of $\mathcal{U}_{m}$. To be more precise, starting from the $\mathcal{U}_{m}$ describing the confinement of parameters $V_{\text{in}} = 2 V_{0}$, $R_{\text{in}} = 0.5 \sqrt{2} l_{B}$, $V_{\text{ext}} = 10 V_{0}$, and $R_{\text{ext}} = 5.7 \sqrt{2} l_{B}$, which we denote by $\mathcal{U}^{(0)}_{m}$, we construct the other sets $\mathcal{U}^{(i)}_{m}$ such that $\mathcal{U}^{(i)}_{m+i} = \mathcal{U}^{(0)}_{m}$ and $\mathcal{U}^{(i)}_{k} \simeq \mathcal{U}^{(0)}_{0}$ for $0 \leq k < m$. As one can check on the plots on the left column of Fig.~\ref{fig:shifted_pot}, this basically reduces to a global shift of the $\mathcal{U}_{m}$ towards higher-$m$ values and to the introduction of some arbitrary potentials for the lowest-$m$ values. This freedom of choice comes from the fact that for all eigenstates of present interest, the occupation of the lowest-$m$ single-particle orbitals is zero.

It is well known from the general theory of the quantum Hall effect~\cite{TongNotes,Goerbig} that QH excitations of a $\nu = 1/2$ FQH liquid are $1/2$ anyons, also called semions. So, if the spectra and/or the density profiles characterizing the EEs were in any way affected by the QH anyonic statistics, one should observe a periodicity in the EE properties distinguishing between cases in which the number of anyons pinned in the origin is even or odd. 

From our calculations, it turns out that this is unfortunately not the case. No periodic behavior as a function of the number of QHs at the center of the droplet can be observed in the energy spectra or in the edge states density profiles, respectively shown in the right column of Fig.~\ref{fig:shifted_pot} and in Fig.~\ref{fig:shifted_dens}. On this basis, neither of these quantities appears to be suitable to point out the QH fractional statistics. An alternative and completely different strategy in this direction has been explored in~\cite{OUetal_TOF}.

Nevertheless, some interesting features can be still seen in the spectra presented in Fig.~\ref{fig:shifted_pot}. For instance, one can note that by increasing the number of QHs in the ground state the energies of the edge states slightly decrease. In particular, the two states described by Jacks with EPs $\eta = \kappa^{n} + \rho$ and $\eta = \kappa^{n+i} + \rho$ end up having different energies even if the $\mathcal{U}_{m}$ values are properly set. This reflects the different length of the edges sustaining a given excitation. As shown by the rescaled density differences $\mathcal{D}(r)$ in Fig.~\ref{fig:shifted_dens}, if such an excitation lies on a longer edge it induces a smaller density deformation; actually the density deformation can be imagined to be globally the same but distributed over a longer region. This provides a reasonable explanation of why it ends up being less affected by the confinement.

\section{\label{sec:conclusions}CONCLUSIONS}

\begin{figure*}
\includegraphics[width=1\textwidth]{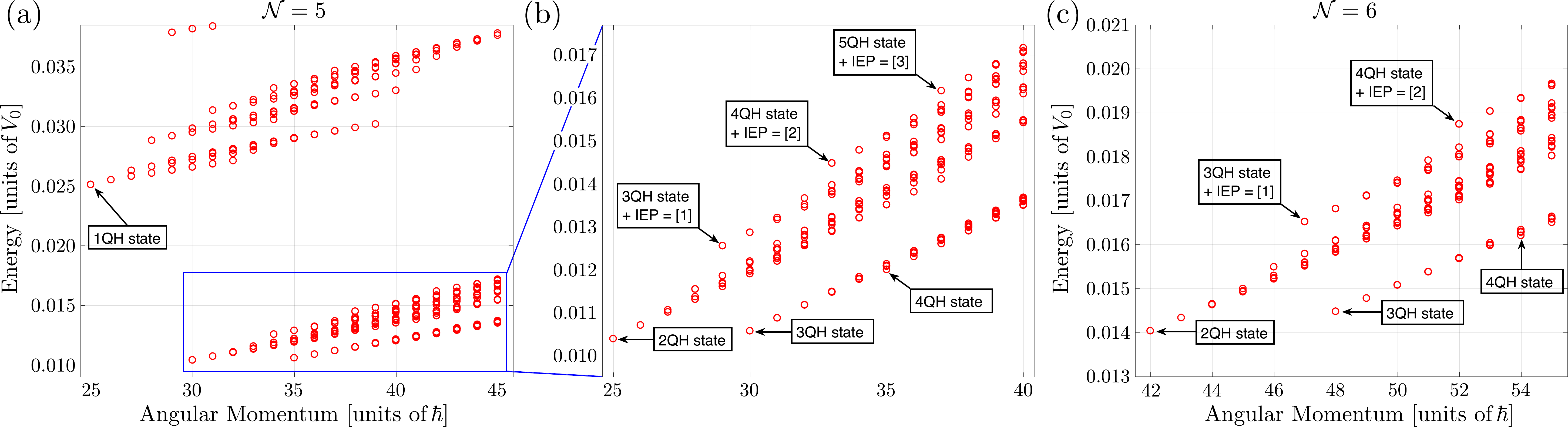}
\caption{Many-body spectra for (a) and (b) $\mathcal{N} = 5$ and (c) $\mathcal{N}=6$ particle systems subject to a piercing potential of parameters $V_{\text{in}} = 2 V_{0}$ and $R_{\text{in}} = 0.5 \sqrt{2} l_{B}$ and a harmonic potential characterized by $\upsilon = 3 \times 10^{-4} V_{0}/ \hbar$. As we can see, the addition of a harmonic confinement preserves the organization of many-body energy levels into branches and subbranches as in the pierced geometry, but at the same time leads to a non-degenerate ground state corresponding to a QH state.}
\label{fig:HWharmonic_spectra}
\end{figure*}

In this work we have made use of numerical exact diagonalization calculations to investigate the properties of trapped $\nu=1/2$ fractional quantum Hall droplets of strongly interacting bosons subject to a synthetic magnetic field and to a pair of concentric, cylindrically symmetric hard-wall potentials which confine the FQH fluid in a ring-shaped region of space bounded by two edges.

Building upon our results for disk-shaped droplets discussed in our previous work~\cite{EMIC_disk}, we have shown how the HW confinement in a ring geometry can be used to obtain a non-degenerate ground state characterized by a ring-shaped density profile and several quasiholes located at the trap center and to lift the degeneracy of edge states with the same total angular momentum. While the presence of two edges complicates the structure of the many-body spectrum, still the eigenstates can be understood in terms of trial wave functions based on Jack polynomials which show excellent overlap with the exact diagonalization eigenstates and offer a convenient way to classify the excited states. Regimes of parameters showing well-defined branches of edge excitations lying on either the inner or the outer edge are identified.

Going beyond this work, these results suggest how ring-shaped geometries obtained by means of HW potentials could be used in a driven-dissipative photonic context to optimize the generation of ring-shaped FQH fluids with an arbitrary number of QHs at the origin. Furthermore, they also represent a unique foundation for future experimental studies of the linear and nonlinear properties of FQH edge excitations in both ultracold atoms and photonic systems. 

From a theoretical perspective, the present work completes the study started in~\cite{EMIC_disk} and provides an exhaustive description of the effects of HW potentials on $\nu=1/2$ bosonic FQH droplets in different geometries. In particular, the proposed extension of the Jack polynomials formalisms, together with a careful analysis of the $\mathcal{U}_{m}$ coefficients describing the external potentials, can be used in the diagonalization procedure to optimize numerical calculations as explained in Appendix~\ref{app:ED}. Finally, the one-to-one correspondence between the system eigenstates and the Jack trial wave functions, which persists independently of the parameters of the hard-wall potentials, seems to indicate a deeper connection between the FQH physics and the Jack polynomial structure. All these directions are presently the subject of on-going studies.

\section{\label{sec:acknowledgments}ACKNOWLEDGMENTS}
This work was supported by the EU-FET Proactive grant AQuS, Project No. 640800, and by the Autonomous Province of Trento, partially through the project ``On silicon chip quantum optics for quantum computing and secure communications" (``SiQuro"). Stimulating discussions with M. Dalmonte, R. Fern, L. Mazza, N. Regnault, and R. Santachiara are warmly acknowledged.

\appendix

\section{\label{app:HW+harmonic} Pierced geometry with a harmonic external confinement}
In this appendix we briefly discuss the situation in which a FQH liquid in the so-called pierced geometry experiences an additional external confinement in the form of a harmonic potential.

If we restrict ourselves to the LLL, the second quantized form of such a confining potential simply gives an energy shift proportional to the total angular momentum
\begin{equation}
\Delta E_{\text{harm}} \propto \upsilon L_{z}
\end{equation}
and does not modify the Hamiltonian eigenstates~\cite{Cazalilla}. Therefore, one can use this scheme to remove the degeneracy between the $n$-QH states and their OEEs lying in the same energy branch, obtaining in this way a non-degenerate ground state with a given number of QHs (see Fig.~\ref{fig:HWharmonic_spectra}). As already mentioned many times in the main text, the number $n$ of QHs in the ground state depends on the parameters of the piercing potential and, in our case, of the strength $\upsilon$ of the external harmonic potential. As a result, even though a quite precise fine-tuning of both of them may be still needed, selecting a precise number of QHs in the ground state with a harmonic confinement is definitely simpler than in the ring geometry where one would have to properly tune four parameters and even small changes of each of them, especially the radii $R_{\text{in}, \text{ext}}$, may strongly affect the ordering of the  eigenstates.

In conclusion, exactly as one can remove the macroscopic degeneracy characterizing an unconfined FQH liquid to obtain a non-degenerate ground state in a Laughlin form, one can add an external harmonic potential to a FQH liquid in the pierced geometry to obtain a non-degenerate ground state in the form of an $n$-QH state. 

On the other hand, the addition of the harmonic confinement preserves the energy branch and subbranch structure distinctive of the pierced geometry (compare Figs.~\ref{fig:singleHW_spectra} and \ref{fig:HWharmonic_spectra}). This means that most of the OEEs of a given $n$-QH state remain very difficult to resolve. As a result, such a confinement of the FQH liquid is the most suitable if one wants to study states containing a different number of QHs but is not suitable to study the excitations lying on their edges.

\section{\label{app:ED} Diagonalization procedure}
In this appendix we present an efficient way to properly choose the basis for the diagonalization procedure of the Hamiltonian describing a $\nu = 1/r$ FQH liquid in the ring geometry. Such a method is mainly based on the following observations: (i) The number of QHs in the ground state can be predicted from the behavior of the $\mathcal{U}_{m}$ potentials as a function of $m$, (ii) the Hamiltonian eigenstates are in general well-described by Jacks~\footnote{The fact that one describes the Hamiltonian eigenstates with Jacks of the form $J^{\alpha}_{\Omega + \eta}$, instead of the more accurate product of Jacks $J^{\nu}_{\eta} J^{\alpha}_{\Omega}$, is completely irrelevant in this context since both of them lie in the same squeezed Hilbert space $\mathcal{H}_{sq}(\Omega+\eta)$.}, and (iii) the squeezed Hilbert spaces associated with partitions $\lambda$ and $\mu$ of the same integer $L = L_{0} + \Delta L$, such that $\lambda \succeq \mu$, are related by \eqref{eq:squeezed_H_relation}.

Once we know the number $n$ of QHs in the system ground state, the monotonically increasing behavior of the $\mathcal{U}_{m}$ with respect to their minimum gives information on the energetic ordering of states containing an increasing or decreasing number of QHs,
\begin{eqnarray}
&E_{n-\text{QH}}& < E_{(n-1)-\text{QH}} < E_{(n-2)-\text{QH}} < \dots  \\ &E_{n-\text{QH}}& < E_{(n+1)-\text{QH}} < E_{(n+2)-\text{QH}} < \dots ,
\end{eqnarray}
but also on the energy costs associated with different EEs sharing the same total angular momentum. In particular, for each angular momentum sector, the lowest-energy eigenstates will be those described by Jacks having root partitions which are not so different from the ground-state one, or rather root partitions which can be obtained from the ground state one by moving particles only by a few orbitals.
So points (i) and (ii) allows us to have a coarse idea of the energetic ordering of Jacks in a given angular momentum sector.

Before moving on, an explanation of what we mean by ``coarse idea of the energetic ordering'' is needed. As we have already discussed, the energy shifts induced by the confining Hamiltonian depend on both the details of the $\mathcal{U}_{m}$ potentials and the particular values of the occupation numbers $n_{m}$ characterizing the different Jacks. As a consequence, the energetic ordering of Jacks with similar root partitions is in general very difficult to predict. For instance, if we consider Jacks of root partitions $\lambda_{1} = \Omega + \kappa^{n} + [1,1] - [1] = \Omega + [n+1,n+1,n \dots, n, n-1]$ and $\lambda_{2} =\Omega + \kappa^{n} +[1,1,1] - [1,1] = \Omega + [n +1, n+1, n+1, n, \dots, n, n-1, n-1]$ they will have almost the same energy and we cannot be sure of which one will have the lowest one. However, without a doubt, both of them will have energies much lower than the one of the Jack polynomial with root partition $\lambda_{3} =\Omega + \kappa^{n-1} + [N+2] = \Omega + [n+N+1, n-1, \dots, n-1]$. The latter indeed has non-vanishing occupation of larger-$m$ single-particle orbitals and, as a consequence, its energy will take contributions from $\mathcal{U}_{m}$ potentials with much higher values. 

At this point the relations between the so-called squeezed Hilbert spaces come into play. Indeed, root partitions obtained from the ground-state one by moving particles only by a few orbitals not only typically label the lowest-energy states, but in general they are also dominated by those labeling the higher-energy ones. In light of the relation \eqref{eq:squeezed_H_relation}, this means that the lowest-energy states of a given angular momentum sector generally lie in a sub-space of the squeezed Hilbert spaces associated with the root partitions of Jacks describing higher-energy states. So, in order to obtain the lowest-energy part of the spectrum, one only has to identify a Jack polynomial of partition $\bar{\lambda}$ with high enough energy (actually one for each angular momentum sector of interest) and diagonalize the Hamiltonian on the squeezed Hilbert space $\mathcal{H}_{\text{sq}} (\bar{\lambda})$.

\begin{figure}[t]
\includegraphics[width=0.4025\textwidth]{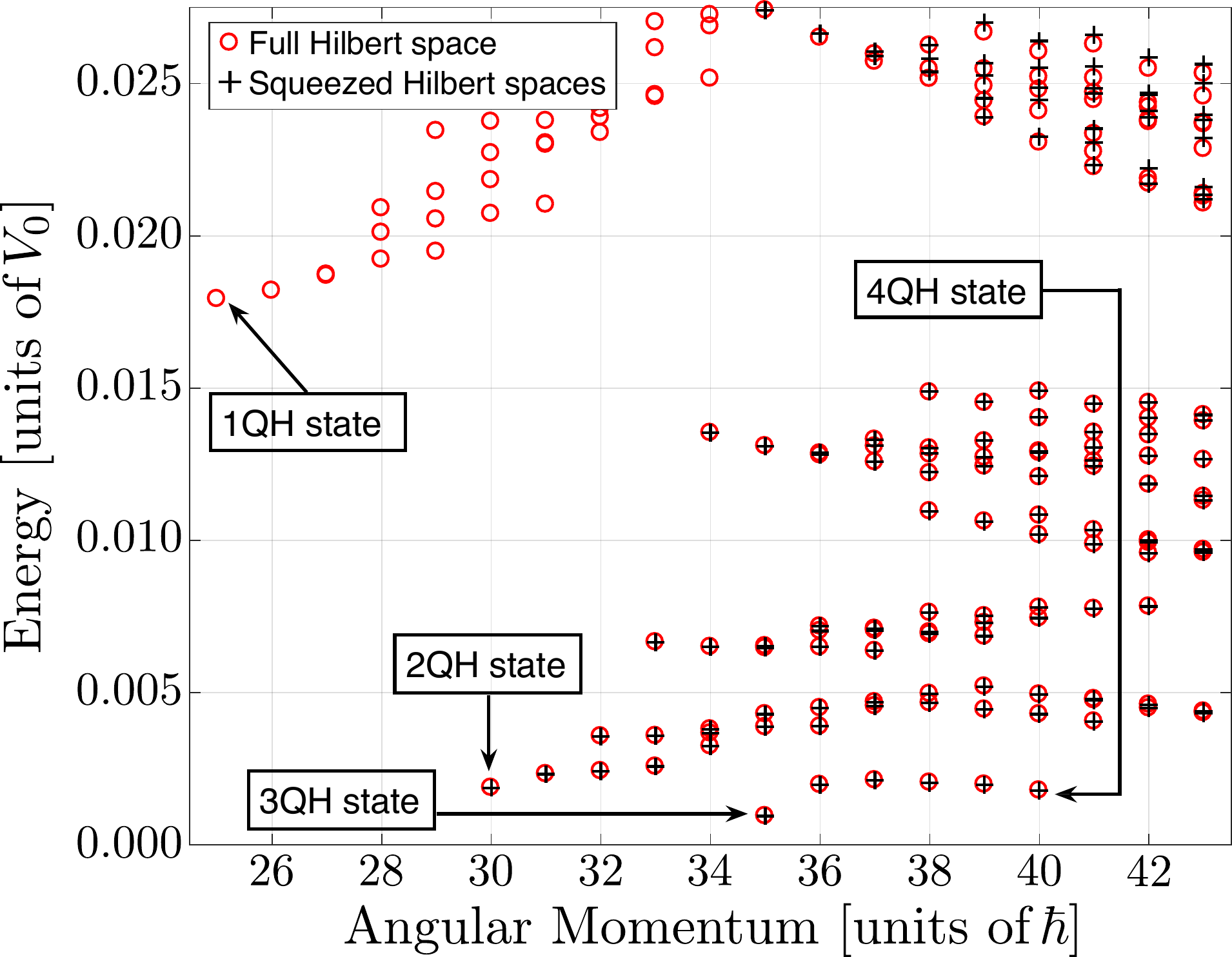}
\caption{Energy spectra for an $\mathcal{N} = 5$ particle system subject to a confinement potential of parameters $V_{\text{in}} = 2 V_{0}$, $R_{\text{in}} = 0.5 \sqrt{2} l_{B}$, $V_{\text{ext}} = 10 V_{0}$, and $R_{\text{ext}} = 5.4 \sqrt{2} l_{B}$. Red circles are obtained by diagonalizing the system Hamiltonian on the full Hilbert space $\mathcal{H}_{\text{full}}$ and black pluses by restricting the system Hamiltonian to the squeezed Hilbert spaces $\mathcal{H}_{\text{sq}} (\bar{\lambda})$. To be precise, for $\mathcal{H}_{\text{full}}$ a single-particle angular momentum cutoff $m=15$ has been chosen, while for the different $\mathcal{H}_{\text{sq}} (\bar{\lambda})$ partitions of the form $\bar{\lambda} = \Omega + \kappa^{2} + \rho$, where $\rho$ is the dominant configuration among those of the integer $L - L_{L} - 2 \mathcal{N}$ which satisfy $\rho_{1} \leq 5$, have been considered. As we can see, the agreement between the different spectra is excellent and small deviations can only be observed for the highest-energy states.}
\label{fig:jacks_vs_fullbasis}
\end{figure}

An example of the efficiency of this method is given in Fig.~\ref{fig:jacks_vs_fullbasis}, where we compare the energy spectra for an $\mathcal{N}=5$ particle system experiencing an external potential of parameters $V_{\text{in}} = 2 V_{0}$, $R_{\text{in}} = 0.5 \sqrt{2} l_{B}$, $V_{\text{ext}} = 10 V_{0}$, and $R_{\text{ext}} = 5.4 \sqrt{2} l_{B}$ obtained by considering the full Hilbert space or a properly chosen squeezed Hilbert spaces.

To be precise, for the full Hilbert space a cutoff on the single-particle angular momenta given by $m=15$ has been chosen. The dimension of such a Hilbert space is $d(\mathcal{H}_{\text{full}}) = 20349$. Regarding the squeezed Hilbert spaces instead, partitions of the form $\bar{\lambda} = \Omega + \kappa^{2} + \rho$ have been considered. In particular, for each angular momentum sector, we have chosen $\rho = [\rho_{1}, \rho_{2}, \dots]$ as the dominant partition among those of the integer $L - L_{L} - 2 \mathcal{N}$ which fulfill the condition we imposed on the maximum single-particle angular momentum for the full Hilbert space. Note that in terms of the elements $\rho$ of the partition, such a constraint corresponds to $\rho_{1} \leq 5$. The sum of the dimensions of the Hilbert spaces obtained in this way is $\sum_{\bar{\lambda}} d(\mathcal{H}_{\text{sq}}(\bar{\lambda})) = 2846$, which is almost one order of magnitude smaller than $d(\mathcal{H}_{\text{full}})$. 

An even smarter choice for the cutoff on the $\rho_{1}$ value could be made by looking at the $\mathcal{U}_{m}$ potentials associated with the considered confining parameter [see Fig.~\ref{fig:HWHW_spectra}(a)]. Since $\mathcal{U}_{1}$ and $\mathcal{U}_{15}$ are of the same order of magnitude and the above choice of $\bar{\lambda}$ neglects states with a non-vanishing occupation of the $m=1$ orbital, the same should be done with those states taking contributions from the $m=15$ orbital by imposing $\rho_{1} \leq 4$. In this way we would correctly reproduce only those states whose energies are lower than $0.015 V_{0}$.

Before concluding let us make some final remarks. First of all, we would like to stress that the relation
\begin{equation}
E(\lambda) > E(\mu) \Longrightarrow \lambda \succeq \mu
\end{equation}
is not true in general, especially when one considers states with total angular momentum quite different from the ground state one. What instead holds is that
\begin{equation}
E(\lambda) \gg E(\mu) \Longrightarrow \lambda \succeq \mu. 
\end{equation}
This means that the choice of $\bar{\lambda}$ must be made in a proper way. Second, it is important to highlight that the approximation introduced by diagonalizing the Hamiltonian on the squeezed Hilbert space $\mathcal{H}_{\text{sq}}(\bar{\lambda})$ is only good as long as the effects of the confinement are weak and mixing with states lying above the Laughlin gap remains negligible. If this condition is violated, the squeezed Hamiltonian $\mathcal{H}_{\text{sq}}(\bar{\lambda})$ may be missing some physical states.


%

\end{document}